\documentclass[sigconf,nonacm=true,natbib=true,balance=false]{acmart}

\usepackage{amsmath}
\usepackage{algorithm}
\usepackage{algpseudocode}
\usepackage{array,multirow,dashrule,colortbl}
\usepackage{multicol}
\algrenewcommand\algorithmicrequire{\textbf{Input:}}
\algrenewcommand\algorithmicensure{\textbf{Output:}}
\algrenewcommand\algorithmiccomment[1]{\hspace{0.5em}\textnormal{// #1}}

\setcopyright{none}
\acmDOI{}
\acmISBN{}

\title{CACO: Coverage-Aware Coordinated Optimization for Continuous Generative Retrieval}

\makeatletter
\newcommand{\noinstitution}{\global\@ACM@instpresenttrue}
\makeatother

\author{Pan Zhang}
\affiliation{%
  \institution{Hunan University}
  \city{Changsha}
  \country{China}
}
\email{hnuzhangpan@hnu.edu.cn}

\author{Nan Zhuang}
\affiliation{%
  \noinstitution
  \city{Beijing}
  \country{China}
}
\email{zhuangn53@pku.edu.cn}

\author{Yongkang Zhang}
\affiliation{%
  \institution{Peking University}
  \city{Beijing}
  \country{China}
}
\email{yongkangzhang273@gmail.com}

\author{Kun X{\'u}}
\affiliation{%
  \noinstitution
  \city{Beijing}
  \country{China}
}
\email{xukunxkxk@gmail.com}

\author{Zhongtao Jiang}
\affiliation{%
  \noinstitution
  \city{Beijing}
  \country{China}
}
\email{ji4ngxiaoming@gmail.com}

\author{Zhiwei Wang}
\affiliation{%
  \noinstitution
  \city{Beijing}
  \country{China}
}
\email{zhiwei.wang1991@gmail.com}

\author{Kun X{\v{u}}}
\authornote{Corresponding authors.}
\affiliation{%
  \noinstitution
  \city{Beijing}
  \country{China}
}
\email{syxu828@gmail.com}

\author{Youhuan Li}
\authornotemark[1]
\affiliation{%
  \institution{Hunan University}
  \city{Changsha}
  \country{China}
}
\email{liyouhuan@hnu.edu.cn}

\renewcommand{\shortauthors}{Pan Zhang et al.}

\begin{document}
\begin{abstract}
Continuous generative retrievers powered by flow matching and diffusion sample multiple queries in item embedding space for approximate nearest-neighbor (ANN) search under fixed budgets.
However, existing generative retrievers can suffer from retrieval overlap under uncoordinated sampling, whereas explicit diversity guidance to broaden the candidate set may induce relevance drift as it lacks direct feedback on the value of additional items.
We introduce \textbf{CACO} (\textbf{C}overage-\textbf{A}ware \textbf{C}oordinated \textbf{O}ptimization), a self-supervised post-training framework coordinating queries from frozen flow-matching and diffusion retrievers toward complementary item coverage with negligible latency overhead.
Using empirical retrieval frequencies across sampled reference queries as a relevance proxy, CACO formulates a label-free coverage objective that improves recall across all 30 backbone--dataset settings in offline experiments.
To avoid the overhead of sampling reference queries at inference, CACO employs a coordinated query adjustment network combining inter-query bidirectional attention with gated relative directions, updating queries via bounded tangent-space projection and spherical retraction.
During training, CACO applies tangential antithetic perturbations and counterfactual single-query replacement to measure each query's marginal gain, yielding an unbiased gradient estimator of smoothed set coverage across non-differentiable ANN search.
On eight public benchmarks and an in-house 24-partition streaming industrial dataset, CACO improves Recall@100 and Recall@400 by up to 23.7\% and 23.9\% over the strongest baseline, cuts inter-query overlap by 53.3\%--64.9\%, and adds only 0.5\%--3.0\% inference latency.

\end{abstract}

\maketitle

\begin{figure}[t]
  \centering
  \includegraphics[width=\linewidth]{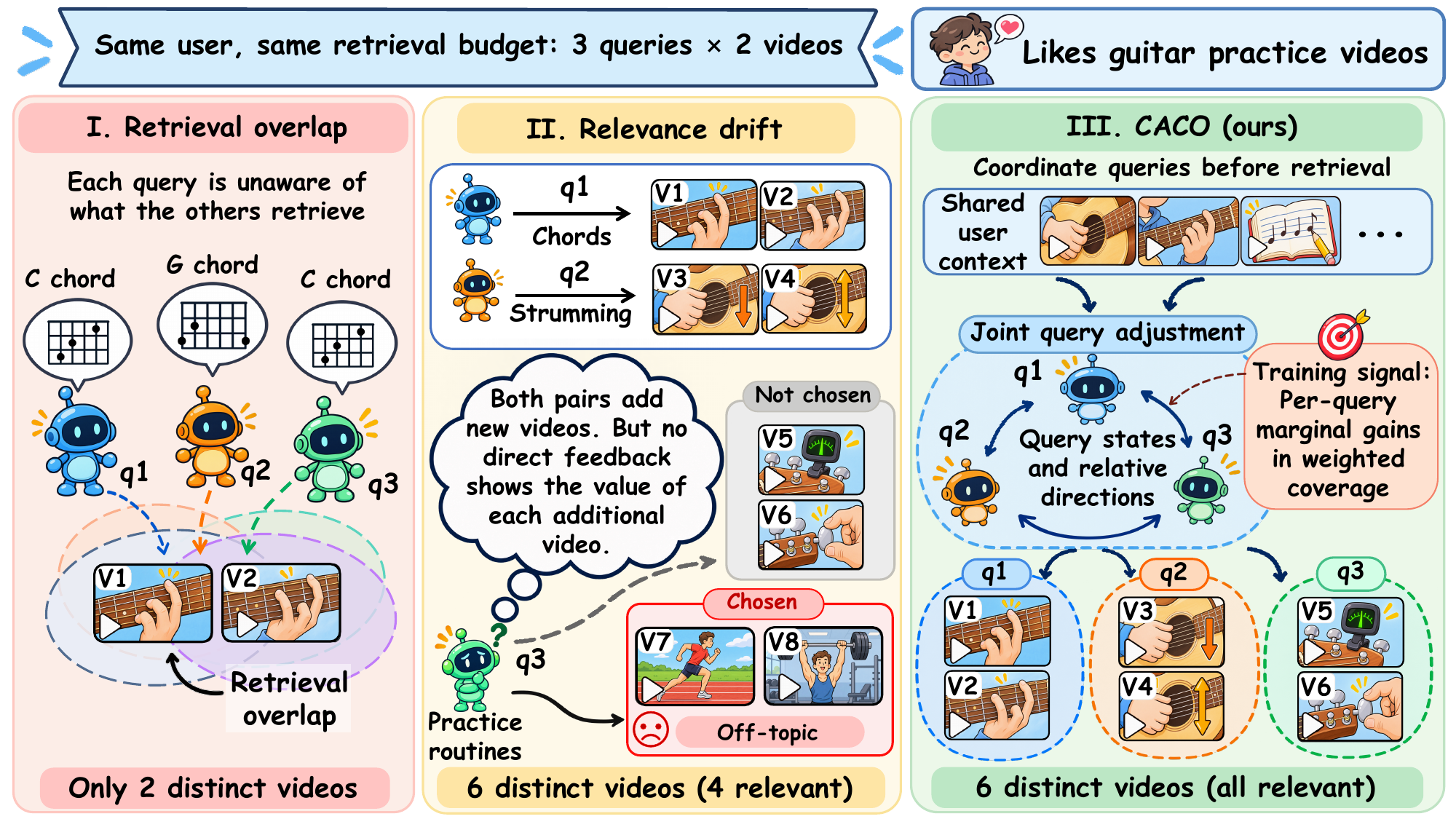}
  \caption{Illustrative retrieval for a user interested in guitar practice, with $L=3$ queries and $K=2$ videos per query. (I) Overlapping results yield only two distinct videos. (II) Six distinct videos are retrieved, but only four are relevant to guitar practice. (III) CACO coordinates the queries to retrieve six distinct, relevant videos.}
  \Description{Three panels show retrieval for the same user under a fixed budget. In Panel I, all three queries retrieve the same two chord videos, V1 and V2. In Panel II, the first two queries retrieve chord and strumming videos V1 through V4. Both options for the third query add two new videos. The practice-routines query retrieves running and weightlifting videos V7 and V8, missing guitar-tuning videos V5 and V6. In Panel III, CACO adjusts the queries jointly using the shared user context, with per-query marginal gains in weighted coverage as the training signal. The three queries retrieve chord, strumming, and tuning videos V1 through V6.}
  \label{fig:coverage-motivation}
\end{figure}

\section{Introduction}
\label{sec:introduction}
\begingroup
\setlength{\emergencystretch}{1em}
Large-scale recommender systems in e-commerce, short-video feeds, and online news have to retrieve a limited set of relevant candidates from massive corpora under strict latency budgets~\cite{covington2016youtube,gao2022kuairec,wu2019nrms,cen2020comirec,li2019mind}.
Recent continuous generative retrievers powered by diffusion~\cite{yang2023dreamrec,li2025dimerec} and flow matching~\cite{chen2024rfm,liu2025fmrec,jin2026confm,yang2026pdflow,shen2026xrec} model user-conditioned query distributions in item embedding space and retrieve candidates via approximate nearest-neighbor (ANN) search~\cite{malkov2020hnsw}.
This generative process allows the system to sample multiple queries under the same user context and merge their retrieval results into a final candidate set, thereby better capturing diverse user interests~\cite{li2025dimerec,cai2025diffdiv}.

However, generating multiple queries alone does not automatically guarantee capturing diverse interests: maximizing recall requires set-level complementarity, where a query adds value only when it retrieves relevant items missed by the others~\cite{agrawal2009diversifying}.
Standard generative retrievers enrich the query distribution across user preferences, yet still sample queries independently at inference~\cite{li2024diffurec,yang2023dreamrec,liu2025preferdiff,mao2025tarec,liu2025fmrec,li2025dimerec}.
Because independent samples remain blind to one another's results, even a diverse distribution can map onto overlapping top-$K$ ANN neighborhoods, causing \emph{retrieval overlap} that wastes limited slots on duplicates.
To break this overlap, recent methods go further by steering queries apart via semantic anchors~\cite{shen2026xrec} or diversity-aware guidance~\cite{cai2025diffdiv}.
Although such guidance broadens the candidate set, it lacks direct feedback from retrieval results on the value of newly added items, risking \emph{relevance drift} that trades duplicate items for off-topic candidates.

Figure~\ref{fig:coverage-motivation} illustrates both failure modes for a user interested in guitar practice under a budget of $L=3$ queries and $K=2$ videos per query.
Uncoordinated queries in Panel~I cluster around chords and repeatedly retrieve the same two videos, returning only two distinct items across six slots.
In contrast, in Panel~II, explicit guidance pushes the third query beyond chords and strumming toward generic practice routines.
Without direct feedback on the value of new results, it picks off-topic fitness videos over guitar tuning, leaving only four of the six videos relevant.
Rather than steering queries directly, an alternative workaround for overlap is to increase the retrieval depth $K$ and deduplicate the results.
Nevertheless, such a compensatory strategy trades extra retrieval overhead for broader coverage, leaving query generation unchanged without optimizing each query's contribution to the merged candidate set.
Achieving the complementary, fully relevant set in Panel~III motivates our central question: under a fixed retrieval budget, how can we coordinate queries so that each retrieves relevant items missed by the others?

In practice, learning to coordinate queries toward this goal presents two challenges.
First, user interaction data provide only partial supervision: indiscriminately expanding coverage may introduce irrelevant items, while fitting a single observed target cannot capture the full range of interests that the query set should cover.
We therefore need a coverage objective that approximates item relevance beyond the single observed target and rewards queries for covering complementary candidates.
Second, even with a suitable coverage objective that can identify high-quality query sets, using it to train query generation remains challenging.
On the one hand, coverage depends on the discrete candidate sets returned by approximate nearest-neighbor search, which prevents direct optimization of query generation through standard backpropagation~\cite{plotz2018n3,xie2020differentiabletopk}.
On the other hand, direct distillation of these high-quality query sets does not necessarily preserve their broader coverage of relevant items, because matching query representations and optimizing set-level retrieval coverage are different objectives.
Preserving this coverage advantage during learning therefore requires training signals derived from actual retrieval results.

To address these challenges, we propose \textbf{CACO} (\textbf{C}overage-\textbf{A}ware \textbf{C}oordinated \textbf{O}ptimization), a lightweight self-supervised post-training framework that coordinates queries from frozen flow-matching and diffusion retrievers.
CACO constructs a label-free, frequency-weighted coverage objective using item retrieval counts across reference queries sampled from the frozen base model as a relevance proxy.
Each covered item contributes its weight only once, so retrieving duplicates yields zero additional gain.
Applying offline greedy selection over a large reference pool confirms that this objective consistently guides query selection toward higher recall without ground-truth target labels.
To avoid generating a large query pool at inference, CACO employs a coordinated query adjustment network that integrates context cross-attention, inter-query bidirectional attention, and gated relative directions, updating queries by bounded tangent-space projection and spherical retraction.
During training, CACO applies tangential antithetic perturbations and holds the other queries fixed via counterfactual single-query replacement to measure each query's marginal coverage gain.
This formulation yields an unbiased gradient estimator of Gaussian-smoothed set coverage across non-differentiable ANN search, improving complementary coverage while keeping the base generator, query count, and retrieval depth unchanged.
\par
\endgroup

Our main contributions are as follows:
\begin{itemize}
    \item \textbf{Label-Free Coverage Objective.} We formulate a fre\-quen\-cy-weighted set-coverage objective that uses empirical retrieval counts across sampled reference queries as a relevance proxy, and verify via offline greedy selection that it improves recall across all 30 backbone--dataset settings.
    \item \textbf{Self-Supervised Post-Training Coordination.} We propose CACO, a lightweight self-supervised post-training framework that coordinates queries sampled from a frozen base generator under a fixed retrieval budget, deriving an unbiased smoothed-coverage gradient estimator from tangential antithetic perturbations and counterfactual single-query replacement.
    \item \textbf{Public and Industrial Validation.} Across eight public benchmarks, an in-house 24-partition streaming industrial dataset, and six frozen flow-matching and diffusion backbones, CACO improves Recall@100 and Recall@400 by up to 23.7\% and 23.9\%, cuts query overlap by 53.3\%--64.9\%, and adds only 0.5\%--3.0\% inference latency.
\end{itemize}

\section{Related Work}
\label{sec:related-work}

\subsection{Generative Retrieval}
\label{sec:related-generative}
\begingroup
\setlength{\emergencystretch}{2em}
Generative retrieval in recommender systems follows discrete or continuous paradigms.
Discrete methods generate semantic item IDs~\mbox{\cite{hou2026generativesurvey}} constructed through vector quantization~\mbox{\cite{rajput2023tiger,hou2023vqrec}} or hierarchical clustering~\mbox{\cite{si2024seater}}.
Later work incorporates collaborative alignment into tokenization and generation~\mbox{\cite{zheng2024lcrec,wang2024letter,liu2025etegrec}} or performs preference alignment~\mbox{\cite{deng2025onerec}}.
Alternative generation schemes include multi-token parallel decoding~\mbox{\cite{hou2025rpg}}, discrete-diffusion parallel decoding in LLaDARec~\mbox{\cite{shi2026lladarec}}, and latent-token-prefixed autoregressive decoding in LATTE~\mbox{\cite{hou2026latte}}.

Continuous methods generate query embeddings conditioned on user context for nearest-neighbor search.
Diffusion-based methods model target representations via iterative denoising~\mbox{\cite{li2024diffurec,yang2023dreamrec}}, quantized condition guidance~\mbox{\cite{mao2025diqdiff}}, or multi-interest and diversity-aware guidance~\mbox{\cite{li2025dimerec,cai2025diffdiv}}.
ADRec~\mbox{\cite{chen2025adrec}} combines autoregressive modeling with item-level diffusion and multi-stage alignment, while PreferDiff~\mbox{\cite{liu2025preferdiff}} formulates Bayesian personalized ranking using generation probabilities.
For efficient generation, TA-Rec~\mbox{\cite{mao2025tarec}} uses consistency regularization for single-step generation, FMRec~\mbox{\cite{liu2025fmrec}} adopts straight-path flow matching---further extended with contrastive alignment~\mbox{\cite{jin2026confm}} and popularity debiasing~\mbox{\cite{yang2026pdflow}}---and X-Rec~\mbox{\cite{shen2026xrec}} combines anchor conditioning with Riemannian flow matching~\mbox{\cite{chen2024rfm}} on the unit sphere.
Order-agnostic continuous item identifiers also enable parallel generation~\mbox{\cite{lin2025setrec}}.
Building on continuous generative retrieval, CACO introduces a self-supervised post-training layer that coordinates queries from frozen flow-matching and diffusion backbones for complementary set-level coverage.

\subsection{Multi-Interest Retrieval}
\label{sec:related-multi-interest}
Multi-interest retrieval represents each user with multiple vectors to capture diverse preference facets for parallel nearest-neighbor search.
Early approaches extract interests from behavior sequences through dynamic capsule routing~\mbox{\cite{li2019mind}} or self-attentive aggregation~\mbox{\cite{cen2020comirec}}.
SINE~\mbox{\cite{tan2021sine}} infers a sparse set of active concepts, while behavior clustering provides another way to construct multiple user representations~\mbox{\cite{pal2020pinnersage}}.
Subsequent work enriches interest modeling with temporal patterns and user profiles~\mbox{\cite{chen2021pimi,chai2022umi}}.

Beyond interest extraction, disagreement and backward-flow regularization promote more informative and diverse representations~\mbox{\cite{li2022miner,zhang2022re4}}, while graph-based item partitioning supports interest disentanglement~\mbox{\cite{du2024dismir}}.
REMI~\mbox{\cite{xie2023remi}} further improves interest learning through interest-aware hard negative mining and routing regularization.
At the supervision level, TiMiRec~\mbox{\cite{wang2022timirec}} introduces target-interest distillation to guide multi-interest learning.
Moving beyond single-positive supervision, MIMA~\mbox{\cite{mao2026mima}} groups concurrent positives within a request, models interest dependencies via a causally masked decoder, and applies Hungarian matching to exclusively assign distinct positive items to different interests.

Similarly, continuous generative retrievers capture diverse preferences through multi-query sampling and explicit diversity guidance~\mbox{\cite{li2025dimerec,cai2025diffdiv,shen2026xrec}}.
However, differentiating representations or guiding queries in isolation from merged retrieval results leaves a coverage gap: independent queries can still retrieve overlapping nearest-neighbor sets, whereas ungrounded query separation risks relevance drift toward off-topic items.
Instead of relying solely on query- or target-level differentiation, CACO coordinates queries directly through marginal gains in frequency-weighted candidate coverage.
\par
\endgroup

\section{Problem Formulation}
\label{sec:problem-formulation}
This section formalizes multi-query continuous generative retrieval under a fixed retrieval budget.
A request is described by context $h=(S_h,A_h,c)$, where $S_h=(v_1,\ldots,v_T)$ is a sequence of historical items ordered by interaction time, $A_h=(a_1,\ldots,a_T)$ contains the corresponding actions, and $c$ denotes additional conditioning information, such as the expected action on the target item~\mbox{\cite{shen2026xrec,liu2024mbgen}}.
Let $\mathcal I$ be the item corpus and $e_v\in\mathbb R^d$ the fixed embedding of item $v\in\mathcal I$.
Given $h$, a model generates a collection of $L$ query vectors:
\begin{equation}
Q=(q_1,\ldots,q_L)\sim\pi_\theta(\cdot\mid h),
\qquad q_i\in\mathbb R^d.
\label{eq:query-distribution}
\end{equation}
Independent conditional sampling is a special case, with $\pi_\theta(Q\mid h)=\prod_{i=1}^L p_\theta(q_i\mid h)$.
Although we use the term query set for convenience, $Q$ is an indexed collection and may contain repeated or similar vectors.

Let $\mathcal A_K(q)\subseteq\mathcal I$ denote the top-$K$ candidate set returned for query $q$ via approximate nearest-neighbor search over $\{e_v\}_{v\in\mathcal I}$.
The final candidate set is the union
\begin{equation}
\mathcal U_K(Q)=\bigcup_{i=1}^L\mathcal A_K(q_i),
\qquad |\mathcal U_K(Q)|\leq LK.
\label{eq:retrieval-union}
\end{equation}
The budget $B=LK$ counts retrieval slots before filtering and deduplication; repeated items do not increase the final candidate count.
Our goal is to learn queries that jointly cover the user's relevant items while keeping $L$ and $K$ fixed.
We evaluate this goal using the next observed positive-interaction item as a sample of the user's latent interests.

\section{Method}
\label{sec:method}
\begingroup
\setlength{\emergencystretch}{1em}
This section presents CACO, our self-supervised post-training framework for multi-query coordination.
We first formulate the label-free coverage objective and verify it via offline greedy selection, and then detail the coordinated query adjustment network and its marginal-coverage training objective.
\par
\endgroup

\subsection{Frequency-Weighted Coverage Objective}
\label{sec:coverage-objective}
\begin{figure}[t]
  \centering
  \includegraphics[trim=174.5bp 188.5bp 357bp 91bp,clip,width=\columnwidth]{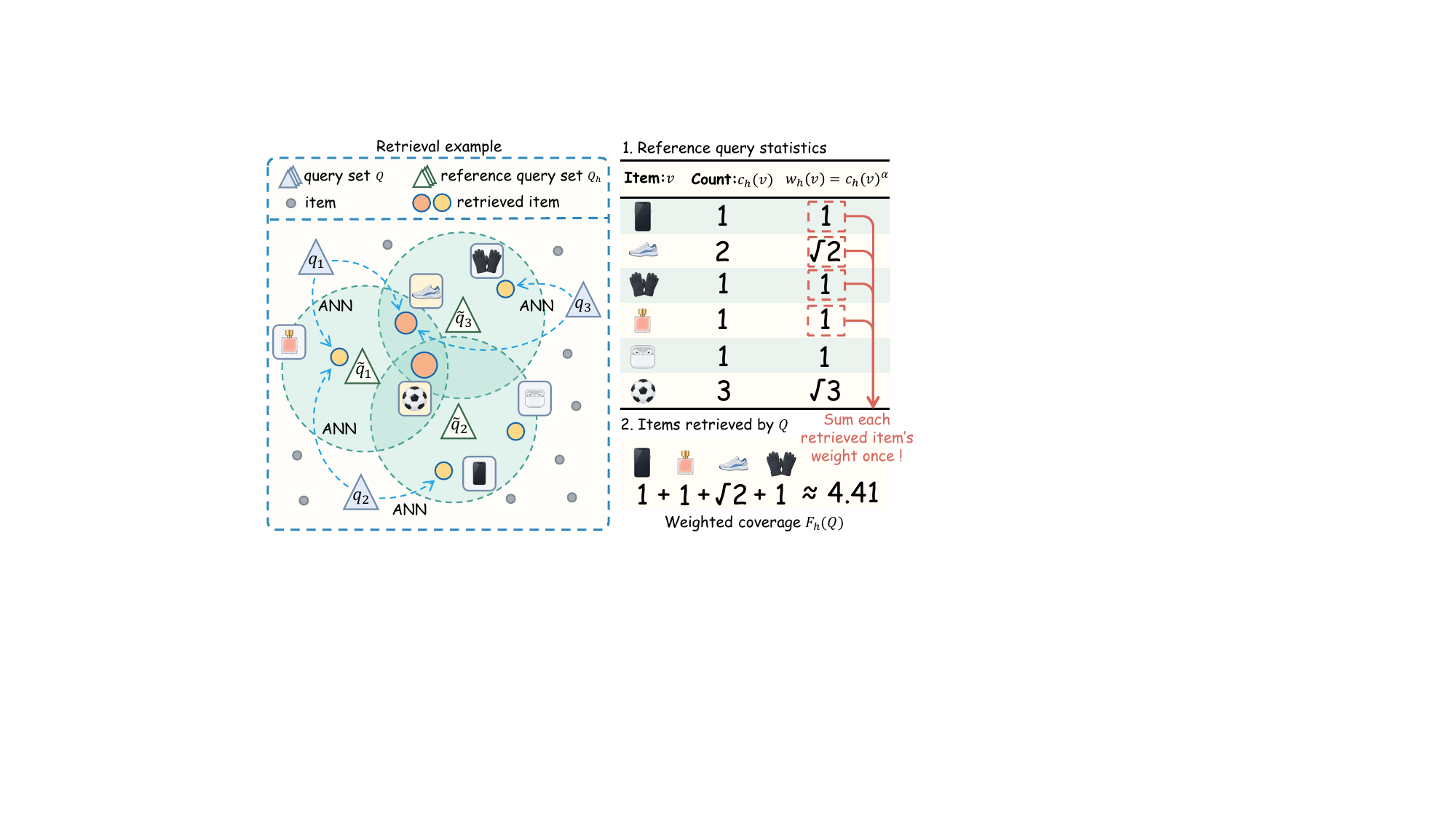}
  \caption{Constructing frequency-weighted coverage. Item weights are derived from retrieval frequencies across the reference queries in $Q_h$. Each unique item retrieved by $Q$ (phone, perfume, shoes, and gloves) contributes once; the illustrated set has total weight $3+\sqrt{2}\approx4.41$ ($\alpha=0.5$).}
  \Description{A retrieval example is paired with reference-query statistics and a coverage calculation. Phone, shoe, gloves, perfume, earbuds, and ball have reference retrieval counts of one, two, one, one, one, and three, respectively. With exponent one half, their weights are the square roots of those counts. The evaluated query set retrieves phone, perfume, shoe, and gloves. Counting each item once gives one plus one plus square root of two plus one, approximately 4.41. This is an illustrative example, not an experimental result.}
  \label{fig:coverage-construction}
\end{figure}

\begin{figure*}[t]
  \centering
  \includegraphics[trim=71bp 72bp 108bp 73.5bp,clip,width=0.9\textwidth]{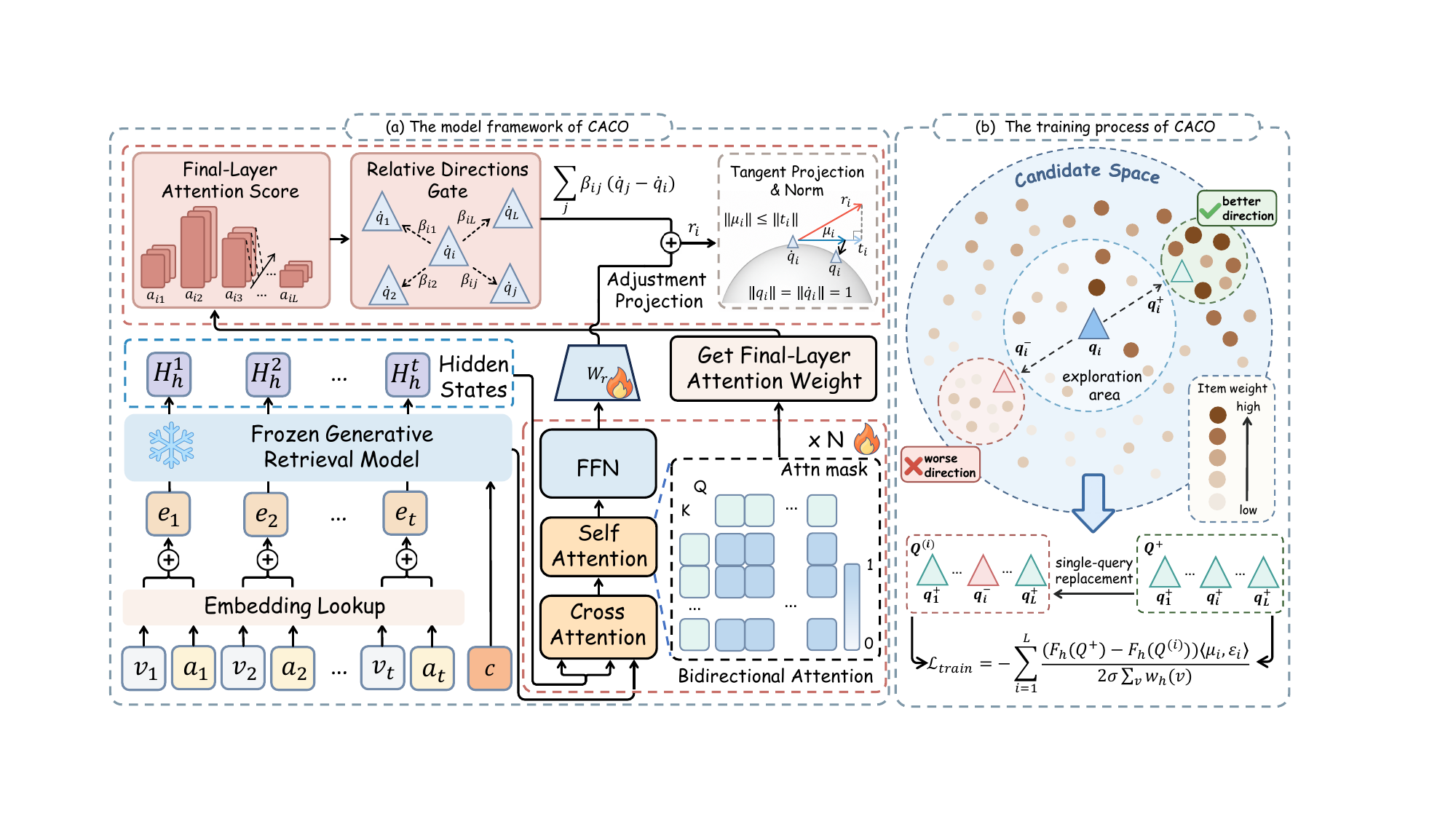}
  \caption{Overview of CACO. (a) The adjustment network uses user history and interactions among queries to predict an adjustment for each query from the frozen generator. It combines this prediction with a learned weighted sum of relative query directions, whose weights are computed from the final query states and the last-layer self-attention weights. The combined adjustment is then projected onto the tangent space, and the final updated query is normalized onto the unit hypersphere. (b) With other queries fixed, training compares opposite perturbations $q_i^+$ and $q_i^-$. The weighted coverage difference guides updates toward high-weight items not already covered by other queries.}
  \Description{The left panel shows a frozen generator and a trainable query adjustment network. Repeated cross-attention, self-attention, and feed-forward blocks produce query states. Final-layer attention weights help determine the weights for aggregating relative query directions. These directions and a predicted residual are combined, projected onto the tangent space, and used to update unit-normalized queries. The right panel illustrates antithetic perturbations, replacement of one positive query by its negative counterpart while retaining the other queries, and the resulting difference in weighted coverage used for training.}
  \label{fig:method-architecture}
  \label{fig:method-overview}
\end{figure*}

An ideal query set covers items in proportion to their relevance to the user.
Let $p^*(v\mid h)$ be the unknown probability that item $v$ is the user's next positive interaction under context $h$.
The probability mass covered by $Q$ is
\begin{equation}
J_h(Q)=\sum_{v\in\mathcal U_K(Q)}p^*(v\mid h).
\label{eq:ideal-coverage}
\end{equation}
Since observed interactions do not reveal the full distribution $p^*(v\mid h)$, we use item retrieval frequencies from the base generator as an empirical relevance proxy under context $h$ (see Appendix~\ref{sec:theoretical-analysis} for its probability-mass interpretation).

For each user context, we independently sample a reference query set $Q_h=(\tilde q_1,\ldots,\tilde q_M)$ from the frozen generator $p_{\theta_0}(\cdot\mid h)$ and collect reference query statistics by computing the retrieval count of each item $v$:
\begin{equation}
c_h(v)=\sum_{m=1}^M\mathbf 1\{v\in\mathcal A_K(\tilde q_m)\}.
\label{eq:item-frequency}
\end{equation}
We use this context-specific count as a relevance proxy and assign item weights
\begin{equation}
w_h(v)=c_h(v)^\alpha,
\qquad \alpha\in(0,1],
\label{eq:frequency-weights}
\end{equation}
with $w_h(v)=\mathbf 1\{c_h(v)>0\}$ when $\alpha=0$.
Notably, setting $\alpha=0$ gives equal weight to all items retrieved by $Q_h$, whereas $\alpha=1$ uses their retrieval counts directly.
Intermediate values (e.g., $\alpha=0.5$ in Figure~\ref{fig:coverage-construction}) retain frequency differences while reducing the dominance of high-frequency items.

The resulting frequency-weighted coverage objective is
\begin{equation}
F_h(Q)=\sum_{v\in\mathcal U_K(Q)}w_h(v).
\label{eq:weighted-coverage}
\end{equation}
Figure~\ref{fig:coverage-construction} illustrates how this objective evaluates a generated query set $Q$ using the item weights derived from the reference query set $Q_h$.
Once an item is covered, retrieving it with another query contributes no additional gain.
A query is therefore valuable when it retrieves high-weight items absent from the other queries' results.
The reference query statistics and weights are computed without ground-truth target item information and remain fixed when comparing query sets.
Items never retrieved by $Q_h$ have $w_h(v)=0$ and contribute zero to $F_h$, but remain eligible for recall evaluation.

\raggedbottom
To verify whether optimizing $F_h(Q)$ improves retrieval recall before training a new model, we first apply an offline greedy selection algorithm over the $M$-query reference query set $Q_h$.
Starting with an empty collection $Q=\emptyset$, the algorithm iteratively selects $L$ queries from $Q_h$.
At each step, it chooses the remaining query with the largest marginal coverage gain:
\begin{equation}
\Delta_h(q\mid Q)=
\sum_{v\in\mathcal A_K(q)\setminus\mathcal U_K(Q)}w_h(v).
\label{eq:greedy-gain}
\end{equation}
After each selection, the covered-item set $\mathcal U_K(Q)$ is updated and the marginal gains of the remaining queries are recomputed.

Section~\ref{sec:greedy-results} shows that, for all tested positive values of $\alpha$, greedy optimization of $F_h(Q)$ improves recall across all 30 backbone--dataset settings.
However, greedy selection requires generating a large query pool ($M\gg L$) and executing nearest-neighbor search for all $M$ queries at inference time, incurring substantial overhead.
To avoid this cost, CACO instead learns to directly adjust a fixed-size collection of $L$ queries.

\subsection{Coordinated Query Adjustment}
\label{sec:query-adjustment}
To adjust queries toward complementary coverage, each query needs access to both the user context and the other queries to account for the items they may retrieve.
Therefore, CACO employs a coordinated query adjustment network to jointly refine the queries, as illustrated in Figure~\ref{fig:method-architecture}(a).
Specifically, the frozen base model provides unit-normalized initial queries $Q^{(0)}=(\dot q_1,\ldots,\dot q_L)$ sampled independently from the generator, along with hidden states $H_h=(H_h^1,\ldots,H_h^T)$ from its history encoder.
Given these inputs, the adjustment network outputs the refined queries $Q=(q_1,\ldots,q_L)$, allowing each adjusted query to depend on the entire initial collection.
Each of the $N$ adjustment layers grounds the queries in the user's history via cross-attention~\mbox{\cite{vaswani2017transformer}} while enabling inter-query coordination via bidirectional self-attention~\mbox{\cite{devlin2019bert}}:
\begin{equation}
\begin{aligned}
S^{(\ell)}&=X^{(\ell-1)}+
\operatorname{CrossAttn}(\overline X^{(\ell-1)},\overline H_h,\overline H_h),\\
U^{(\ell)}&=S^{(\ell)}+
\operatorname{SelfAttn}(\overline S^{(\ell)}),\\
X^{(\ell)}&=U^{(\ell)}+
\operatorname{FFN}(\overline U^{(\ell)}),
\end{aligned}
\label{eq:interaction-layers}
\end{equation}
where $X^{(0)}$ denotes the projected initial queries, $\ell=1,\ldots,N$, and the overbar denotes layer normalization.
In addition to predicting each query's adjustment from its final hidden state, we use a relative directions gate to add a weighted sum of directions toward the other queries:
\begin{equation}
r_i=W_r\overline X_i^{(N)}+
\sum_{j=1}^L\beta_{ij}(\dot q_j-\dot q_i).
\label{eq:relative-residual}
\end{equation}
The gating weights $\beta_{ij}$ are computed from the final query states and the last-layer self-attention weights $a_{ij}^{(a)}$ across $A$ attention heads:
\begin{equation}
\beta_{ij}=\frac{1}{\sqrt A}\sum_{a=1}^A
\bigl(w_a^\top\overline X_i^{(N)}\bigr)a_{ij}^{(a)},
\label{eq:relative-gates}
\end{equation}
where $w_a$ is a learnable vector shared across queries in the same attention head.
Allowing $\beta_{ij}$ to take either sign lets each query move toward or away from the others, paralleling signed gates that learn context-dependent directions from paired counterfactuals~\mbox{\cite{li2026same}}.

Following prior work on spherical retraction~\mbox{\cite{absil2012retractions,bonnabel2013rsgd,boumal2023manifolds}}, we project $r_i$ onto the tangent space at $\dot q_i$ via $P_i(v)=v-\langle v,\dot q_i\rangle\dot q_i$ and apply a bounded update:
\begin{equation}
t_i=P_i(r_i),\qquad
\mu_i=\frac{t_i}{\sqrt{1+\|t_i\|^2/\rho^2}},\qquad
q_i=\frac{\dot q_i+\mu_i}{\sqrt{1+\|\mu_i\|^2}},
\label{eq:bounded-update}
\end{equation}
where $\rho>0$ is a hyperparameter that bounds the step size.
Since $\mu_i\perp\dot q_i$ and $\|\mu_i\|<\rho$, the updated query $q_i$ preserves unit norm while restricting its deterministic angular displacement to at most $\arctan\rho$.
Zero-initializing the residual and gating output layers ensures that the adjustment network starts from an identity mapping ($q_i=\dot q_i$), preserving the initial queries at the beginning of training while keeping the query count $L$ and retrieval depth $K$ unchanged.

\subsection{Training via Marginal Coverage Gains}
\label{sec:coverage-learning}
Training the adjustment network to maximize $F_h(Q)$ is non-trivial because the discrete nearest-neighbor retrieval sets in $F_h(Q)$ prevent direct backpropagation.
To overcome this, CACO constructs a surrogate training signal by comparing positive and negative perturbations of each query while holding the other queries' retrieval results fixed, as illustrated in Figure~\ref{fig:method-architecture}(b).
Specifically, given an initial query collection $Q^{(0)}$ sampled independently of the reference query set $Q_h$, the adjustment network produces the bounded tangential step $\mu_i$ for each query $i$ as in Eq.~\eqref{eq:bounded-update}.
To explore the local neighborhood around the predicted direction, we independently sample Gaussian noise $\epsilon_i\sim\mathcal N(0,I)$ for each query and perturb $\mu_i$ in the tangent space to form an antithetic query pair~\mbox{\cite{salimans2017es,mania2018ars}}:
\begin{equation}
q_i^\pm=\frac{\dot q_i+\mu_i^\pm}{\sqrt{1+\|\mu_i^\pm\|^2}},
\qquad
\mu_i^\pm=\mu_i\pm\sigma P_i(\epsilon_i),
\label{eq:perturbed-queries}
\end{equation}
where $\sigma>0$ is the perturbation scale controlling the local exploration area around $q_i$.

Let $Q^+=(q_1^+,\ldots,q_L^+)$ denote the positively perturbed collection.
For each query $i$, we construct a counterfactual collection $Q^{(i)}$ via single-query replacement, substituting only $q_i^+$ in $Q^+$ with $q_i^-$.
We then measure the normalized coverage difference:
\begin{equation}
\Delta_i=\frac{F_h(Q^+)-F_h(Q^{(i)})}
{\sum_{v\in\mathcal I}w_h(v)}.
\label{eq:counterfactual-gain}
\end{equation}
Since all other queries remain unchanged, $\Delta_i$ directly compares the marginal coverage gains of the two perturbations.
Changes that affect only items already covered by the other queries yield zero gain, whereas newly covered items contribute their frequency weights, making the signal sensitive to both estimated relevance and what the other queries retrieve.

Finally, we optimize the adjustment network by aligning the tangential update $\mu_i$ with the coverage-improving perturbation direction:
\begin{equation}
\mathcal L_{\mathrm{train}}=
-\sum_{i=1}^L\frac{\Delta_i
\langle\mu_i,\epsilon_i\rangle}{2\sigma},
\label{eq:coverage-loss}
\end{equation}
where $\langle\cdot,\cdot\rangle$ denotes the inner product.
A positive $\Delta_i$ pulls $\mu_i$ toward $+\epsilon_i$ (the better direction in Figure~\ref{fig:method-architecture}(b)), whereas a negative $\Delta_i$ pushes it toward $-\epsilon_i$.
We prove in Appendix~\ref{sec:theoretical-analysis} (Theorem~\ref{thm:unbiased-coverage-gradient}) that, with fixed reference weights, the negative gradient of $\mathcal L_{\mathrm{train}}$ is an unbiased estimator of the gradient of the Gaussian-smoothed normalized coverage objective.
During training, only the adjustment network is updated while the base generator and item embeddings remain frozen.
Because the positive and negative retrieval sets can be reused across the per-query comparisons, each training step requires only two retrievals per query (i.e., $2L$ per query collection) rather than re-running retrieval for every perturbed collection.

Furthermore, because the quality of the training signal $\Delta_i$ depends on the reference query set $Q_h$, we can refresh $Q_h$ and the item weights $w_h(v)$ during training using the adjusted queries produced by CACO---which are more dispersed and may uncover potentially relevant items missed by the base generator---to provide stronger supervision for subsequent training.

\section{Experiments}
\label{sec:experiments}
\raggedbottom
\begin{table*}[t]
\caption{Overall retrieval performance on eight public datasets (Recall on a 0--1 scale; higher is better).}
\label{tab:overall-retrieval}
\centering
\definecolor{CovACShade}{HTML}{FBEEDE}
\definecolor{CovACInk}{HTML}{000000}
\newcommand{\SecondBestResult}[1]{\begingroup\setbox0=\hbox{#1}\leavevmode\rlap{\raisebox{-1.4pt}[0pt][0pt]{\hdashrule{\wd0}{0.35pt}{1.4pt 1pt}}}\box0\endgroup}
\fontsize{8.5}{10.625}\selectfont
\setlength{\tabcolsep}{0pt}
\renewcommand{\arraystretch}{1.07}
\begin{tabular}{@{}p{18mm}@{\hspace{1.5mm}}p{19mm}@{\hspace{1.5mm}}>{\centering\arraybackslash}p{\dimexpr(\linewidth-50.5mm)/16\relax}>{\centering\arraybackslash}p{\dimexpr(\linewidth-50.5mm)/16\relax}@{\hspace{1.5mm}}>{\centering\arraybackslash}p{\dimexpr(\linewidth-50.5mm)/16\relax}>{\centering\arraybackslash}p{\dimexpr(\linewidth-50.5mm)/16\relax}@{\hspace{1.5mm}}>{\centering\arraybackslash}p{\dimexpr(\linewidth-50.5mm)/16\relax}>{\centering\arraybackslash}p{\dimexpr(\linewidth-50.5mm)/16\relax}@{\hspace{1.5mm}}>{\centering\arraybackslash}p{\dimexpr(\linewidth-50.5mm)/16\relax}>{\centering\arraybackslash}p{\dimexpr(\linewidth-50.5mm)/16\relax}@{\hspace{1.5mm}}>{\centering\arraybackslash}p{\dimexpr(\linewidth-50.5mm)/16\relax}>{\centering\arraybackslash}p{\dimexpr(\linewidth-50.5mm)/16\relax}@{\hspace{1.5mm}}>{\centering\arraybackslash}p{\dimexpr(\linewidth-50.5mm)/16\relax}>{\centering\arraybackslash}p{\dimexpr(\linewidth-50.5mm)/16\relax}@{\hspace{1.5mm}}>{\centering\arraybackslash}p{\dimexpr(\linewidth-50.5mm)/16\relax}>{\centering\arraybackslash}p{\dimexpr(\linewidth-50.5mm)/16\relax}@{\hspace{1.5mm}}>{\centering\arraybackslash}p{\dimexpr(\linewidth-50.5mm)/16\relax}>{\centering\arraybackslash}p{\dimexpr(\linewidth-50.5mm)/16\relax}@{}}
\toprule
 & \textbf{Method} & \multicolumn{2}{c}{\textbf{Beauty}} & \multicolumn{2}{c}{\textbf{Sports}} & \multicolumn{2}{c}{\textbf{Toys}} & \multicolumn{2}{c}{\textbf{Steam}} & \multicolumn{2}{c}{\textbf{Douban}} & \multicolumn{2}{c}{\textbf{Instruments}} & \multicolumn{2}{c}{\textbf{Games}} & \multicolumn{2}{c@{}}{\textbf{Movies}} \\
 & {\fontsize{6.8}{8.4}\selectfont\textit{Users / Items}} & \multicolumn{2}{c}{\fontsize{6.8}{8.4}\selectfont 22.4K / 12.1K} & \multicolumn{2}{c}{\fontsize{6.8}{8.4}\selectfont 35.6K / 18.4K} & \multicolumn{2}{c}{\fontsize{6.8}{8.4}\selectfont 19.4K / 11.9K} & \multicolumn{2}{c}{\fontsize{6.8}{8.4}\selectfont 281K / 12.0K} & \multicolumn{2}{c}{\fontsize{6.8}{8.4}\selectfont 23.8K / 12.6K} & \multicolumn{2}{c}{\fontsize{6.8}{8.4}\selectfont 57.4K / 24.6K} & \multicolumn{2}{c}{\fontsize{6.8}{8.4}\selectfont 94.8K / 25.6K} & \multicolumn{2}{c@{}}{\fontsize{6.8}{8.4}\selectfont 657K / 198K} \\
\cmidrule(lr){3-4}\cmidrule(lr){5-6}\cmidrule(lr){7-8}\cmidrule(lr){9-10}\cmidrule(lr){11-12}\cmidrule(lr){13-14}\cmidrule(lr){15-16}\cmidrule(lr){17-18}
 & & {\fontsize{7.2}{9}\selectfont R@100} & {\fontsize{7.2}{9}\selectfont R@400} & {\fontsize{7.2}{9}\selectfont R@100} & {\fontsize{7.2}{9}\selectfont R@400} & {\fontsize{7.2}{9}\selectfont R@100} & {\fontsize{7.2}{9}\selectfont R@400} & {\fontsize{7.2}{9}\selectfont R@100} & {\fontsize{7.2}{9}\selectfont R@400} & {\fontsize{7.2}{9}\selectfont R@100} & {\fontsize{7.2}{9}\selectfont R@400} & {\fontsize{7.2}{9}\selectfont R@100} & {\fontsize{7.2}{9}\selectfont R@400} & {\fontsize{7.2}{9}\selectfont R@100} & {\fontsize{7.2}{9}\selectfont R@400} & {\fontsize{7.2}{9}\selectfont R@100} & {\fontsize{7.2}{9}\selectfont R@400} \\
\midrule
\multirow{3}{18mm}{\centering\fontsize{7.8}{9.36}\selectfont\shortstack{Multi-Interest\\Retrieval}} & ComiRec & 0.078 & 0.125 & 0.045 & 0.076 & 0.091 & 0.138 & 0.022 & 0.029 & 0.133 & 0.241 & 0.030 & 0.051 & 0.050 & 0.082 & 0.058 & 0.088 \\
 & REMI & 0.067 & 0.096 & 0.039 & 0.058 & 0.063 & 0.086 & 0.099 & 0.141 & 0.255 & 0.379 & 0.050 & 0.075 & 0.073 & 0.111 & 0.044 & 0.062 \\
 & MIMA & 0.099 & 0.157 & 0.050 & 0.083 & 0.096 & 0.150 & 0.113 & 0.207 & 0.225 & 0.389 & 0.064 & 0.101 & 0.115 & 0.177 & 0.049 & 0.073 \\
\midrule
\multirow{3}{18mm}{\centering\fontsize{7.8}{9.36}\selectfont\shortstack{Discrete\\Generative\\Retrieval}} & TIGER & 0.102 & 0.180 & 0.059 & 0.118 & 0.094 & 0.179 & 0.187 & 0.346 & 0.305 & 0.446 & 0.095 & 0.162 & 0.140 & 0.245 & 0.082 & 0.142 \\
 & LATTE & 0.106 & 0.184 & 0.058 & 0.121 & 0.093 & 0.178 & 0.191 & 0.350 & 0.306 & 0.448 & \SecondBestResult{0.096} & \SecondBestResult{0.164} & 0.140 & \SecondBestResult{0.245} & 0.083 & 0.142 \\
 & LLaDARec & 0.053 & 0.092 & 0.037 & 0.063 & 0.045 & 0.085 & \textbf{0.199} & \SecondBestResult{0.361} & 0.297 & 0.442 & 0.069 & 0.102 & 0.099 & 0.168 & 0.083 & 0.139 \\
\midrule
\multirow{8}{18mm}{\centering\fontsize{7.8}{9.36}\selectfont\shortstack{Continuous\\Generative\\Retrieval}} & DreamRec & 0.055 & 0.082 & 0.024 & 0.043 & 0.050 & 0.080 & 0.057 & 0.099 & 0.159 & 0.236 & 0.046 & 0.073 & 0.063 & 0.099 & 0.040 & 0.064 \\
 & DimeRec & 0.035 & 0.061 & 0.020 & 0.037 & 0.069 & 0.103 & 0.168 & 0.339 & 0.187 & 0.352 & 0.041 & 0.066 & 0.089 & 0.142 & 0.064 & 0.119 \\
 & PreferDiff & 0.050 & 0.077 & 0.022 & 0.037 & 0.053 & 0.081 & 0.013 & 0.031 & 0.068 & 0.106 & 0.033 & 0.052 & 0.032 & 0.053 & 0.017 & 0.034 \\
 & TA-Rec & 0.064 & 0.096 & 0.029 & 0.049 & 0.061 & 0.092 & 0.017 & 0.031 & 0.057 & 0.100 & 0.030 & 0.049 & 0.048 & 0.080 & 0.012 & 0.022 \\
 & ADRec & 0.065 & 0.095 & 0.037 & 0.058 & 0.063 & 0.091 & 0.079 & 0.125 & 0.213 & 0.293 & 0.053 & 0.078 & 0.082 & 0.122 & 0.045 & 0.067 \\
 & X-Rec & \SecondBestResult{0.115} & \SecondBestResult{0.206} & 0.055 & 0.127 & \SecondBestResult{0.117} & \SecondBestResult{0.207} & 0.133 & 0.317 & 0.242 & 0.449 & 0.058 & 0.138 & 0.096 & 0.217 & 0.058 & 0.119 \\
 & Base & 0.113 & 0.185 & \SecondBestResult{0.074} & \SecondBestResult{0.141} & 0.117 & 0.192 & 0.180 & 0.336 & \SecondBestResult{0.326} & \SecondBestResult{0.473} & 0.088 & 0.159 & \SecondBestResult{0.146} & 0.243 & \SecondBestResult{0.093} & \SecondBestResult{0.154} \\
 & \multicolumn{1}{@{}>{\columncolor{CovACShade}[1mm][\dimexpr1.5mm+0.5pt\relax]}p{19mm}@{\hspace{1.5mm}}}{\textcolor{CovACInk}{\textbf{CACO}}} & \multicolumn{1}{@{}>{\columncolor{CovACShade}[0pt][\dimexpr 0pt+0.5pt\relax]\centering\arraybackslash}p{\dimexpr(\linewidth-50.5mm)/16\relax}@{}}{\textcolor{CovACInk}{\textbf{0.142}}} & \multicolumn{1}{@{}>{\columncolor{CovACShade}[0pt][\dimexpr 1.5mm+0.5pt\relax]\centering\arraybackslash}p{\dimexpr(\linewidth-50.5mm)/16\relax}@{\hspace{1.5mm}}}{\textcolor{CovACInk}{\textbf{0.236}}} & \multicolumn{1}{@{}>{\columncolor{CovACShade}[0pt][\dimexpr 0pt+0.5pt\relax]\centering\arraybackslash}p{\dimexpr(\linewidth-50.5mm)/16\relax}@{}}{\textcolor{CovACInk}{\textbf{0.084}}} & \multicolumn{1}{@{}>{\columncolor{CovACShade}[0pt][\dimexpr 1.5mm+0.5pt\relax]\centering\arraybackslash}p{\dimexpr(\linewidth-50.5mm)/16\relax}@{\hspace{1.5mm}}}{\textcolor{CovACInk}{\textbf{0.175}}} & \multicolumn{1}{@{}>{\columncolor{CovACShade}[0pt][\dimexpr 0pt+0.5pt\relax]\centering\arraybackslash}p{\dimexpr(\linewidth-50.5mm)/16\relax}@{}}{\textcolor{CovACInk}{\textbf{0.141}}} & \multicolumn{1}{@{}>{\columncolor{CovACShade}[0pt][\dimexpr 1.5mm+0.5pt\relax]\centering\arraybackslash}p{\dimexpr(\linewidth-50.5mm)/16\relax}@{\hspace{1.5mm}}}{\textcolor{CovACInk}{\textbf{0.244}}} & \multicolumn{1}{@{}>{\columncolor{CovACShade}[0pt][\dimexpr 0pt+0.5pt\relax]\centering\arraybackslash}p{\dimexpr(\linewidth-50.5mm)/16\relax}@{}}{\textcolor{CovACInk}{\SecondBestResult{0.196}}} & \multicolumn{1}{@{}>{\columncolor{CovACShade}[0pt][\dimexpr 1.5mm+0.5pt\relax]\centering\arraybackslash}p{\dimexpr(\linewidth-50.5mm)/16\relax}@{\hspace{1.5mm}}}{\textcolor{CovACInk}{\textbf{0.393}}} & \multicolumn{1}{@{}>{\columncolor{CovACShade}[0pt][\dimexpr 0pt+0.5pt\relax]\centering\arraybackslash}p{\dimexpr(\linewidth-50.5mm)/16\relax}@{}}{\textcolor{CovACInk}{\textbf{0.357}}} & \multicolumn{1}{@{}>{\columncolor{CovACShade}[0pt][\dimexpr 1.5mm+0.5pt\relax]\centering\arraybackslash}p{\dimexpr(\linewidth-50.5mm)/16\relax}@{\hspace{1.5mm}}}{\textcolor{CovACInk}{\textbf{0.534}}} & \multicolumn{1}{@{}>{\columncolor{CovACShade}[0pt][\dimexpr 0pt+0.5pt\relax]\centering\arraybackslash}p{\dimexpr(\linewidth-50.5mm)/16\relax}@{}}{\textcolor{CovACInk}{\textbf{0.103}}} & \multicolumn{1}{@{}>{\columncolor{CovACShade}[0pt][\dimexpr 1.5mm+0.5pt\relax]\centering\arraybackslash}p{\dimexpr(\linewidth-50.5mm)/16\relax}@{\hspace{1.5mm}}}{\textcolor{CovACInk}{\textbf{0.194}}} & \multicolumn{1}{@{}>{\columncolor{CovACShade}[0pt][\dimexpr 0pt+0.5pt\relax]\centering\arraybackslash}p{\dimexpr(\linewidth-50.5mm)/16\relax}@{}}{\textcolor{CovACInk}{\textbf{0.167}}} & \multicolumn{1}{@{}>{\columncolor{CovACShade}[0pt][\dimexpr 1.5mm+0.5pt\relax]\centering\arraybackslash}p{\dimexpr(\linewidth-50.5mm)/16\relax}@{\hspace{1.5mm}}}{\textcolor{CovACInk}{\textbf{0.301}}} & \multicolumn{1}{@{}>{\columncolor{CovACShade}[0pt][\dimexpr 0pt+0.5pt\relax]\centering\arraybackslash}p{\dimexpr(\linewidth-50.5mm)/16\relax}@{}}{\textcolor{CovACInk}{\textbf{0.104}}} & \multicolumn{1}{@{}>{\columncolor{CovACShade}[0pt][0pt]\centering\arraybackslash}p{\dimexpr(\linewidth-50.5mm)/16\relax}@{}}{\textcolor{CovACInk}{\textbf{0.180}}} \\
\midrule
 & \textit{Improve} & {\fontsize{7.2}{9}\selectfont 23.72\%} & {\fontsize{7.2}{9}\selectfont 14.78\%} & {\fontsize{7.2}{9}\selectfont 13.61\%} & {\fontsize{7.2}{9}\selectfont 23.94\%} & {\fontsize{7.2}{9}\selectfont 20.59\%} & {\fontsize{7.2}{9}\selectfont 17.92\%} & {\fontsize{7.2}{9}\selectfont \textemdash} & {\fontsize{7.2}{9}\selectfont 8.85\%} & {\fontsize{7.2}{9}\selectfont 9.45\%} & {\fontsize{7.2}{9}\selectfont 12.71\%} & {\fontsize{7.2}{9}\selectfont 7.98\%} & {\fontsize{7.2}{9}\selectfont 18.28\%} & {\fontsize{7.2}{9}\selectfont 13.73\%} & {\fontsize{7.2}{9}\selectfont 22.85\%} & {\fontsize{7.2}{9}\selectfont 11.27\%} & {\fontsize{7.2}{9}\selectfont 16.51\%} \\
\bottomrule
\end{tabular}
\par\vspace{4pt}
\begin{minipage}{\linewidth}
\fontsize{8}{10}\selectfont
\microtypesetup{protrusion=false}
\textbf{Bold}/\SecondBestResult{dashed underline}: best/second-best unrounded mean.\\
R@100/R@400 use $10\times10$/$20\times20$ queries--neighbors.\\
For CACO, R@100 is evaluated at $10\times10$ using the adjustment network trained with 20 queries.\\
Improve: positive relative gain over the best other method (including Base). \textemdash: no gain.
\end{minipage}
\end{table*}

\begin{figure*}[t]
  \centering
  \includegraphics[trim=12bp 16bp 10bp 15bp,clip,width=\textwidth]{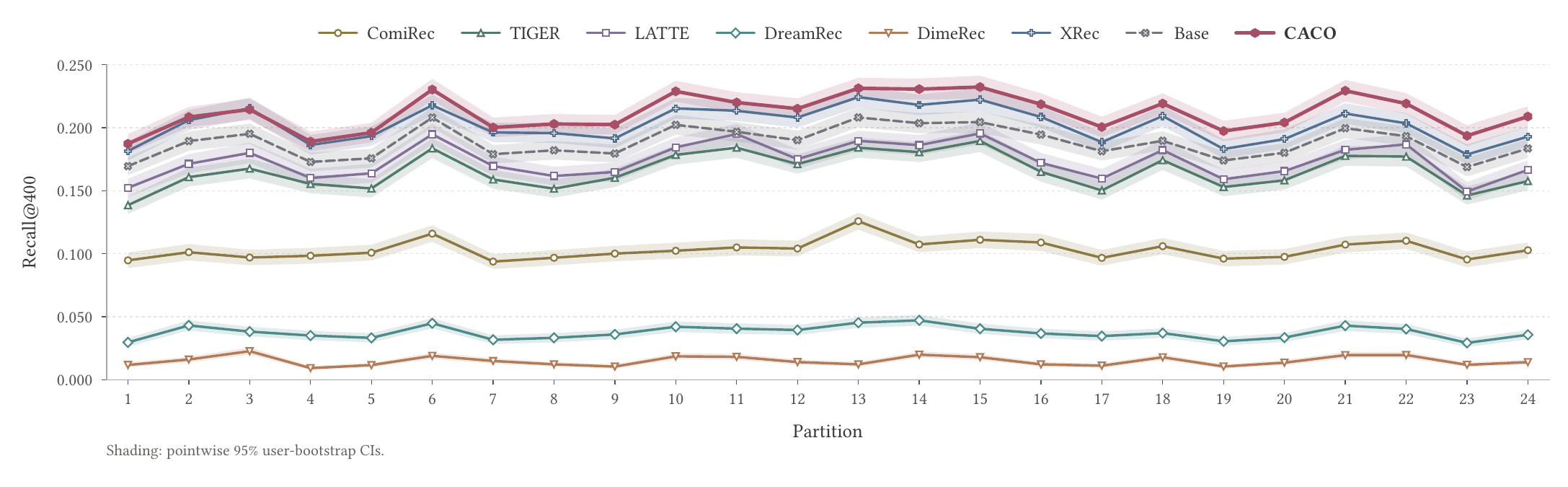}
  \caption{Recall@400 trajectories across 24 consecutive partitions on the in-house industrial dataset under the $20\times20$ retrieval budget. Each point evaluates 10,240 requests. Shaded regions show pointwise 95\% confidence intervals estimated by cluster bootstrap for Recall@400.}
  \Description{Recall curves for eight retrieval methods across 24 consecutive partitions on the in-house industrial dataset. Shaded bands show pointwise 95 percent user-bootstrap confidence intervals, and distinct markers identify the methods. CACO exceeds Base in every partition and has the highest recall in all but the third partition.}
  \label{fig:industrial-streaming}
\end{figure*}

\begin{figure}[t]
  \centering
  \includegraphics[width=\columnwidth]{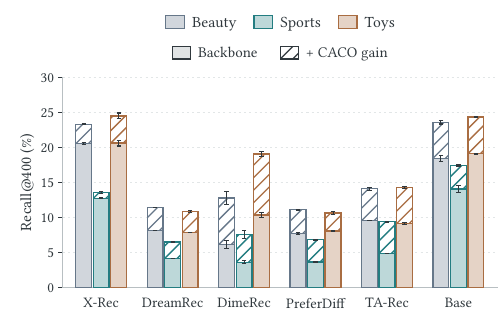}
  \caption{Recall@400 across six frozen generative backbones on Amazon2014 ($20\times20$ budget). Each group shows Beauty, Sports, and Toys from left to right; solid bars show backbone recall and hatched increments show the CACO gain. Error bars show one standard deviation across three training seeds, each averaged over two inference seeds.}
  \Description{A grouped bar chart compares X-Rec, DreamRec, DimeRec, PreferDiff, TA-Rec, and Base across Beauty, Sports, and Toys. Solid lower bars denote frozen backbone recall and diagonally hatched increments denote the CACO gain. All eighteen paired means increase, by 0.89 to 8.72 percentage points.}
  \label{fig:cross-backbone-caco}
\end{figure}

We evaluate CACO across eight public benchmarks, an in-house 24-partition streaming industrial dataset, and six frozen generative backbones, examining overall recall, cross-backbone generalization, overlap reduction and inference efficiency, component ablations, and our frequency-weighted coverage objective.

\subsection{Experimental Setup}
\label{sec:experimental-setup}
\begingroup
\setlength{\emergencystretch}{1em}
\paragraph{Datasets.}
We evaluate on eight public benchmarks and an in-house industrial dataset.
The public benchmarks include Beauty, Sports, and Toys from Amazon Reviews 2014~\mbox{\cite{mcauley2015amazon}}, Musical Instruments (Instruments), Video Games (Games), and Movies and TV (Movies) from Amazon Reviews 2023~\mbox{\cite{hou2026bridging}}, Steam game reviews~\mbox{\cite{kang2018sasrec}}, and Douban Movie from DGRec~\mbox{\cite{song2019dgrec}}. Table~\ref{tab:overall-retrieval} lists user and item counts.
We use the official 5-core subsets for all Amazon datasets and treat reviews as positive implicit feedback~\mbox{\cite{kang2018sasrec}}.
We deduplicate user--item interactions on Steam and Douban. We apply a 5-core filter to both users and games on Steam and retain users with at least five distinct movies on Douban.
Following the standard leave-one-out protocol~\mbox{\cite{kang2018sasrec}}, each user's interactions are ordered by time (with a maximum sequence length of 200), holding out the final item for testing, the penultimate item for validation, and earlier items for training.
For industrial evaluation, we construct an in-house dataset from chronological user--item interaction logs across 24 sequential partitions.
Each partition contains $\approx$2.6M training and 10,240 test instances pairing historical positive and recent interaction sequences, whose target items form that partition's ANN candidate pool.
Due to data security and commercial confidentiality requirements, we keep this in-house dataset proprietary and omit platform-specific details.
\par
\endgroup

\paragraph{Baselines.}
We compare CACO against three categories of representative retrieval methods:
(1) \textbf{Multi-interest retrieval methods}, which learn multiple user interest vectors for parallel nearest-neighbor search, including ComiRec~\mbox{\cite{cen2020comirec}}, REMI~\mbox{\cite{xie2023remi}}, and MIMA~\mbox{\cite{mao2026mima}};
(2) \textbf{Discrete generative retrieval methods}, which tokenize items into semantic IDs and generate target identifiers via sequence decoding or discrete diffusion, including TIGER~\mbox{\cite{rajput2023tiger}}, LATTE~\mbox{\cite{hou2026latte}}, and LLaDARec~\mbox{\cite{shi2026lladarec}}; and
(3) \textbf{Continuous generative retrieval methods}, which generate query embeddings in the continuous item space via diffusion or flow matching for nearest-neighbor retrieval, including DreamRec~\mbox{\cite{yang2023dreamrec}}, DimeRec~\mbox{\cite{li2025dimerec}}, PreferDiff~\mbox{\cite{liu2025preferdiff}}, TA-Rec~\mbox{\cite{mao2025tarec}}, ADRec~\mbox{\cite{chen2025adrec}}, and X-Rec~\mbox{\cite{shen2026xrec}}.
In addition, we include \textbf{Base}, our backbone continuous generator that adopts the sequence architecture of X-Rec~\mbox{\cite{shen2026xrec}} without anchor conditioning and learns query embeddings via Riemannian flow matching on the unit sphere~\mbox{\cite{chen2024rfm}}.
As shown in Figure~\ref{fig:method-architecture}(a), Base constructs each historical interaction token representation by summing its item embedding and action embedding.
Applying CACO on top of the frozen Base model provides a direct comparison to isolate the gain of coordinated query adjustment.

\begingroup
\setlength{\emergencystretch}{3em}
\paragraph{Training.}
Across all compared methods, we use a shared set of fixed, unit-normalized 128-dimensional item embeddings on each dataset.
For public datasets, the item embeddings are pretrained on training sequences using a 2-layer causal Transformer with a next-item prediction objective; for the industrial dataset, each item is represented by a 128-dimensional unit-normalized multimodal embedding.
The public Base generator uses 8 Transformer layers with hidden dimension 256 ($\sim$8M parameters), 30 Euler integration steps, and a classifier-free guidance scale of 3.0, while the adjustment network of CACO sets the number of layers $N=2$, hidden dimension 128, and the number of attention heads $A=4$ ($\sim$0.6M parameters); overall, CACO and all baselines maintain a comparable parameter scale.
On public datasets, all retrieval models are trained for two epochs using AdamW (learning rate $10^{-4}$, batch size 512).
For CACO, we train the adjustment network for the first epoch using the reference query set and item weights from Base, and refresh them using queries adjusted by the first-epoch network before training for the second epoch.
We set $M=200$, $\alpha=0.5$, step bound $\rho=\tan 32^\circ$, and perturbation scale $\sigma=\tan 8^\circ/\sqrt{127}$.

For the industrial dataset, we pretrain each method for 100,000 steps with a batch size of 2,048 on an independent historical corpus, followed by sequential supervised fine-tuning and evaluation over 24 consecutive partitions.
\par
\endgroup

\paragraph{Evaluation protocol and metrics.}
We report $\operatorname{Recall}@B$, the fraction of test targets contained in the merged candidate set $\mathcal U_K(Q)$ (Eq.~\eqref{eq:retrieval-union}), under budgets $L\times K=10\times10$ ($B=100$) and $20\times20$ ($B=400$).
To quantify inter-query redundancy, we define the mean pairwise Jaccard overlap of a query set $Q$ as
\begingroup
\predisplaypenalty=0
\begin{equation}
\operatorname{MPJ}(Q)=\frac{2}{L(L-1)}
\sum_{1\leq i<j\leq L}
\frac{|\mathcal A_K(q_i)\cap\mathcal A_K(q_j)|}
{|\mathcal A_K(q_i)\cup\mathcal A_K(q_j)|}.
\label{eq:jaccard-overlap}
\end{equation}
\endgroup
We average $\operatorname{MPJ}(Q)$ over test contexts; lower values indicate less redundant retrieval.
For all methods, we filter out items that have already appeared in the user's interaction history and deduplicate retrieved candidates.
For the main comparison, we report mean full-test recall across three training seeds (42, 44, 45), each evaluated with inference seeds 42 and 43.

\subsection{Overall Retrieval Performance}
\label{sec:overall-results}
Table~\ref{tab:overall-retrieval} compares all methods on eight public datasets under $10\times10$ ($\operatorname{Recall}@100$) and $20\times20$ ($\operatorname{Recall}@400$) retrieval budgets.
Multi-interest and continuous methods retrieve $K$ nearest neighbors per generated query.
Since directly decoding $B$ candidates entails costly wide-beam search~\mbox{\cite{hou2025rpg,hou2026latte,shen2026xrec}}, we decode $L$ semantic IDs for each discrete generative baseline and expand each decoded item with $K$ nearest neighbors in the shared embedding space, thereby maintaining a unified $L\times K$ retrieval budget.

CACO achieves the highest recall in 15 of 16 dataset--budget settings, with relative gains of up to 23.72\% in $\operatorname{Recall}@100$ and 23.94\% in $\operatorname{Recall}@400$ over the strongest competing method.
The sole exception is Steam at $\operatorname{Recall}@100$, where LLaDARec reaches 0.199 against CACO's 0.196, whereas CACO overtakes it at $\operatorname{Recall}@400$ with 0.393 versus 0.361. On Douban, Base is the strongest competitor at both budgets, and CACO improves on it by 9.45\% and 12.71\%, respectively.
Compared with the frozen Base model sharing the same generator and item embeddings, CACO improves $\operatorname{Recall}@100$ by 1.01--3.08 percentage points and $\operatorname{Recall}@400$ by 2.55--6.02 percentage points without increasing $L$ or $K$.
Absolute gains over Base grow under $20\times20$ where coordination matters more, while training once at $L=20$ transfers directly to $L=10$ at inference.

\subsection{Performance on Industrial Dataset}
\label{sec:industrial-results}
Figure~\ref{fig:industrial-streaming} reports $\operatorname{Recall}@400$ under the $20\times20$ retrieval budget.
CACO achieves the highest overall recall across the streaming partitions, maintaining a 2--3 percentage-point lead over Base throughout evaluation.
It matches X-Rec on early partitions and pulls ahead on later ones, showing that coordinated query adjustment remains effective as user preferences and candidate pools evolve.
More broadly, the flow-matching retrievers CACO, X-Rec, and Base consistently form the strongest group, followed by the discrete generative methods LATTE and TIGER, while ComiRec and multi-step diffusion retrievers remain lower.
CACO thus sustains its gains beyond static public benchmarks on temporally evolving industrial streams.

\begin{table}[!t]
\caption{Retrieval overlap and end-to-end inference cost.}
\label{tab:jaccard-overlap}
\label{tab:overlap-inference-cost}
\begingroup\centering\fontsize{8.5}{10.5}\selectfont
\setlength{\tabcolsep}{0pt}\renewcommand{\arraystretch}{1.12}
\begin{tabular}{@{}p{14mm}@{\hspace{1mm}}>{\centering\arraybackslash}p{\dimexpr(\linewidth-18.3mm)/8\relax}>{\centering\arraybackslash}p{\dimexpr(\linewidth-18.3mm)/8\relax}@{\hspace{1.1mm}}>{\centering\arraybackslash}p{\dimexpr(\linewidth-18.3mm)/8\relax}>{\centering\arraybackslash}p{\dimexpr(\linewidth-18.3mm)/8\relax}@{\hspace{1.1mm}}>{\centering\arraybackslash}p{\dimexpr(\linewidth-18.3mm)/8\relax}>{\centering\arraybackslash}p{\dimexpr(\linewidth-18.3mm)/8\relax}@{\hspace{1.1mm}}>{\centering\arraybackslash}p{\dimexpr(\linewidth-18.3mm)/8\relax}>{\centering\arraybackslash}p{\dimexpr(\linewidth-18.3mm)/8\relax}@{}}
\toprule
\multirow{2}{*}{\textbf{Method}} & \multicolumn{2}{c@{\hspace{1.1mm}}}{\textbf{Beauty}} & \multicolumn{2}{c@{\hspace{1.1mm}}}{\textbf{Sports}} & \multicolumn{2}{c@{\hspace{1.1mm}}}{\textbf{Toys}} & \multicolumn{2}{c}{\textbf{Industrial}} \\
\cmidrule(l{1pt}r{\dimexpr1.1mm+1pt\relax}){2-3}\cmidrule(l{1pt}r{\dimexpr1.1mm+1pt\relax}){4-5}\cmidrule(l{1pt}r{\dimexpr1.1mm+1pt\relax}){6-7}\cmidrule(l{1pt}r{1pt}){8-9}
 & {\fontsize{7.3}{9}\selectfont MPJ\,$\downarrow$} & {\fontsize{7.3}{9}\selectfont Time\,$\downarrow$} & {\fontsize{7.3}{9}\selectfont MPJ\,$\downarrow$} & {\fontsize{7.3}{9}\selectfont Time\,$\downarrow$} & {\fontsize{7.3}{9}\selectfont MPJ\,$\downarrow$} & {\fontsize{7.3}{9}\selectfont Time\,$\downarrow$} & {\fontsize{7.3}{9}\selectfont MPJ\,$\downarrow$} & {\fontsize{7.3}{9}\selectfont Time\,$\downarrow$} \\
\midrule
Base & 0.237 & \textbf{4.322} & 0.185 & \textbf{4.297} & 0.196 & \textbf{4.318} & 0.125 & \textbf{6.454} \\
\textbf{CACO} & \textbf{0.106} & 4.406 & \textbf{0.065} & 4.425 & \textbf{0.081} & 4.426 & \textbf{0.058} & 6.487 \\
\addlinespace[1pt]
\textit{Overhead} &  & {\fontsize{7.2}{9}\selectfont +1.95\%} &  & {\fontsize{7.2}{9}\selectfont +2.97\%} &  & {\fontsize{7.2}{9}\selectfont +2.51\%} &  & {\fontsize{7.2}{9}\selectfont +0.51\%} \\
\bottomrule\end{tabular}
\par\vspace{4pt}\raggedright\fontsize{7.3}{9}\selectfont
MPJ: mean pairwise Jaccard. Time: end-to-end inference time (ms/sample). Industrial reports partition averages.\\
Overhead: percentage increase in end-to-end inference time relative to Base.
\par
\endgroup
\end{table}

\begin{table*}[t]
\caption{Offline query selection across five backbones and six datasets: Recall@400 ($\uparrow$) and $\operatorname{MPJ}$ ($\downarrow$, parentheses; Eq.~\eqref{eq:jaccard-overlap}).}
\label{tab:public-greedy-table}
\definecolor{GreedyShade}{HTML}{FBEEDE}
\definecolor{RandomRatio}{HTML}{276347}
\definecolor{GreedyRatio}{HTML}{A12C43}
\providecommand{\Second}[1]{\begingroup\setbox0=\hbox{#1}\leavevmode\rlap{\raisebox{-1.4pt}[0pt][0pt]{\hdashrule{\wd0}{0.35pt}{1.4pt 1pt}}}\box0\endgroup}
\providecommand{\DatasetHeader}[1]{\makebox[22pt][c]{\textbf{#1}}\hspace{1.4pt}\makebox[34pt][r]{}}
\providecommand{\WithRandomRatio}[2]{\makebox[22pt][c]{#1}\hspace{1.4pt}\makebox[34pt][r]{\textcolor{RandomRatio}{\fontsize{7.5}{9}\selectfont\bfseries (#2)}}}
\providecommand{\WithGreedyRatio}[2]{\makebox[22pt][c]{#1}\hspace{1.4pt}\makebox[34pt][r]{\textcolor{GreedyRatio}{\fontsize{7.5}{9}\selectfont\bfseries (#2)}}}
\begingroup
\centering
\fontsize{8.7}{10.1}\selectfont
\renewcommand{\arraystretch}{1.01}
\setlength{\tabcolsep}{0pt}
\begin{tabular}{@{}p{18mm}p{27mm}*{6}{>{\raggedleft\arraybackslash}p{\dimexpr(\linewidth-45mm)/6\relax}}@{}}
\toprule
\textbf{Backbone} & \textbf{Selection} & \DatasetHeader{Beauty} & \DatasetHeader{Sports} & \DatasetHeader{Toys} & \DatasetHeader{Steam} & \DatasetHeader{Douban} & \DatasetHeader{Industrial} \\
\midrule
\multirow{4}{*}{LATTE} & Random & \WithRandomRatio{0.176}{0.0332} & \WithRandomRatio{0.110}{0.0192} & \WithRandomRatio{0.131}{0.0115} & \WithRandomRatio{0.350}{0.0375} & \WithRandomRatio{0.480}{0.0545} & \WithRandomRatio{0.098}{0.0333} \\
 & Greedy ($\alpha=0$) & \WithGreedyRatio{0.118}{0.0001} & \WithGreedyRatio{0.072}{0.0000} & \WithGreedyRatio{0.091}{0.0000} & \WithGreedyRatio{0.392}{0.0038} & \WithGreedyRatio{0.507}{0.0038} & \WithGreedyRatio{0.091}{0.0064} \\
 & \cellcolor{GreedyShade}Greedy ($\alpha=0.5$) & \cellcolor{GreedyShade}\WithGreedyRatio{\Second{0.200}}{0.0042} & \cellcolor{GreedyShade}\WithGreedyRatio{\Second{0.131}}{0.0042} & \cellcolor{GreedyShade}\WithGreedyRatio{\Second{0.165}}{0.0033} & \cellcolor{GreedyShade}\WithGreedyRatio{\Second{0.424}}{0.0094} & \cellcolor{GreedyShade}\WithGreedyRatio{\Second{0.539}}{0.0092} & \cellcolor{GreedyShade}\WithGreedyRatio{\Second{0.142}}{0.0075} \\
 & Greedy ($\alpha=1$) & \WithGreedyRatio{\textbf{0.218}}{0.0125} & \WithGreedyRatio{\textbf{0.142}}{0.0108} & \WithGreedyRatio{\textbf{0.179}}{0.0087} & \WithGreedyRatio{\textbf{0.426}}{0.0154} & \WithGreedyRatio{\textbf{0.554}}{0.0163} & \WithGreedyRatio{\textbf{0.144}}{0.0087} \\
\noalign{\vskip2pt\hbox{\color[HTML]{666666}\hdashrule{\linewidth}{0.5pt}{4pt 2pt}}\vskip2pt}
\multirow{4}{*}{X-Rec} & Random & \WithRandomRatio{0.205}{0.1119} & \WithRandomRatio{0.127}{0.0371} & \WithRandomRatio{0.204}{0.1316} & \WithRandomRatio{0.316}{0.0193} & \WithRandomRatio{0.449}{0.0324} & \WithRandomRatio{0.197}{0.0042} \\
 & Greedy ($\alpha=0$) & \WithGreedyRatio{0.224}{0.0419} & \WithGreedyRatio{0.105}{0.0043} & \WithGreedyRatio{0.234}{0.0644} & \WithGreedyRatio{0.225}{0.0000} & \WithGreedyRatio{0.335}{0.0000} & \WithGreedyRatio{0.187}{0.0095} \\
 & \cellcolor{GreedyShade}Greedy ($\alpha=0.5$) & \cellcolor{GreedyShade}\WithGreedyRatio{\textbf{0.240}}{0.0503} & \cellcolor{GreedyShade}\WithGreedyRatio{\Second{0.150}}{0.0114} & \cellcolor{GreedyShade}\WithGreedyRatio{\textbf{0.239}}{0.0745} & \cellcolor{GreedyShade}\WithGreedyRatio{\Second{0.393}}{0.0043} & \cellcolor{GreedyShade}\WithGreedyRatio{\Second{0.526}}{0.0029} & \cellcolor{GreedyShade}\WithGreedyRatio{\Second{0.217}}{0.0128} \\
 & Greedy ($\alpha=1$) & \WithGreedyRatio{\Second{0.239}}{0.0614} & \WithGreedyRatio{\textbf{0.155}}{0.0194} & \WithGreedyRatio{\Second{0.237}}{0.0860} & \WithGreedyRatio{\textbf{0.413}}{0.0091} & \WithGreedyRatio{\textbf{0.547}}{0.0066} & \WithGreedyRatio{\textbf{0.218}}{0.0191} \\
\noalign{\vskip2pt\hbox{\color[HTML]{666666}\hdashrule{\linewidth}{0.5pt}{4pt 2pt}}\vskip2pt}
\multirow{4}{*}{DimeRec} & Random & \WithRandomRatio{0.061}{0.6220} & \WithRandomRatio{0.034}{0.6675} & \WithRandomRatio{0.107}{0.6535} & \WithRandomRatio{0.339}{0.0566} & \WithRandomRatio{0.351}{0.0764} & \WithRandomRatio{0.014}{0.0257} \\
 & Greedy ($\alpha=0$) & \WithGreedyRatio{\textbf{0.079}}{0.5393} & \WithGreedyRatio{\Second{0.042}}{0.5998} & \WithGreedyRatio{\textbf{0.126}}{0.5813} & \WithGreedyRatio{0.313}{0.0035} & \WithGreedyRatio{0.296}{0.0025} & \WithGreedyRatio{0.015}{0.0024} \\
 & \cellcolor{GreedyShade}Greedy ($\alpha=0.5$) & \cellcolor{GreedyShade}\WithGreedyRatio{\Second{0.078}}{0.5405} & \cellcolor{GreedyShade}\WithGreedyRatio{\textbf{0.042}}{0.5986} & \cellcolor{GreedyShade}\WithGreedyRatio{\Second{0.125}}{0.5801} & \cellcolor{GreedyShade}\WithGreedyRatio{\Second{0.398}}{0.0121} & \cellcolor{GreedyShade}\WithGreedyRatio{\Second{0.412}}{0.0101} & \cellcolor{GreedyShade}\WithGreedyRatio{\textbf{0.018}}{0.0085} \\
 & Greedy ($\alpha=1$) & \WithGreedyRatio{0.078}{0.5419} & \WithGreedyRatio{0.042}{0.5987} & \WithGreedyRatio{0.125}{0.5802} & \WithGreedyRatio{\textbf{0.406}}{0.0238} & \WithGreedyRatio{\textbf{0.423}}{0.0209} & \WithGreedyRatio{\Second{0.018}}{0.0161} \\
\noalign{\vskip2pt\hbox{\color[HTML]{666666}\hdashrule{\linewidth}{0.5pt}{4pt 2pt}}\vskip2pt}
\multirow{4}{*}{Base} & Random & \WithRandomRatio{0.183}{0.2373} & \WithRandomRatio{0.136}{0.1844} & \WithRandomRatio{0.192}{0.1967} & \WithRandomRatio{0.336}{0.1047} & \WithRandomRatio{0.473}{0.2303} & \WithRandomRatio{0.181}{0.1250} \\
 & Greedy ($\alpha=0$) & \WithGreedyRatio{\Second{0.214}}{0.1402} & \WithGreedyRatio{\Second{0.167}}{0.0616} & \WithGreedyRatio{\Second{0.225}}{0.1191} & \WithGreedyRatio{0.393}{0.0118} & \WithGreedyRatio{0.544}{0.0313} & \WithGreedyRatio{0.200}{0.0339} \\
 & \cellcolor{GreedyShade}Greedy ($\alpha=0.5$) & \cellcolor{GreedyShade}\WithGreedyRatio{\textbf{0.216}}{0.1489} & \cellcolor{GreedyShade}\WithGreedyRatio{\textbf{0.170}}{0.0721} & \cellcolor{GreedyShade}\WithGreedyRatio{\textbf{0.227}}{0.1296} & \cellcolor{GreedyShade}\WithGreedyRatio{\textbf{0.429}}{0.0205} & \cellcolor{GreedyShade}\WithGreedyRatio{\textbf{0.565}}{0.0363} & \cellcolor{GreedyShade}\WithGreedyRatio{\textbf{0.222}}{0.0392} \\
 & Greedy ($\alpha=1$) & \WithGreedyRatio{0.213}{0.1601} & \WithGreedyRatio{0.167}{0.0866} & \WithGreedyRatio{0.224}{0.1414} & \WithGreedyRatio{\Second{0.424}}{0.0337} & \WithGreedyRatio{\Second{0.565}}{0.0457} & \WithGreedyRatio{\Second{0.221}}{0.0493} \\
\noalign{\vskip2pt\hbox{\color[HTML]{666666}\hdashrule{\linewidth}{0.5pt}{4pt 2pt}}\vskip2pt}
\multirow{4}{*}{\textbf{CACO}} & Random & \WithRandomRatio{0.224}{0.1108} & \WithRandomRatio{0.165}{0.0715} & \WithRandomRatio{0.231}{0.0864} & \WithRandomRatio{0.368}{0.0387} & \WithRandomRatio{0.517}{0.0889} & \WithRandomRatio{0.194}{0.0635} \\
 & Greedy ($\alpha=0$) & \WithGreedyRatio{\Second{0.255}}{0.0527} & \WithGreedyRatio{0.183}{0.0253} & \WithGreedyRatio{0.258}{0.0387} & \WithGreedyRatio{0.408}{0.0053} & \WithGreedyRatio{0.538}{0.0099} & \WithGreedyRatio{0.217}{0.0130} \\
 & \cellcolor{GreedyShade}Greedy ($\alpha=0.5$) & \cellcolor{GreedyShade}\WithGreedyRatio{\textbf{0.258}}{0.0627} & \cellcolor{GreedyShade}\WithGreedyRatio{\textbf{0.190}}{0.0355} & \cellcolor{GreedyShade}\WithGreedyRatio{\textbf{0.265}}{0.0494} & \cellcolor{GreedyShade}\WithGreedyRatio{\textbf{0.451}}{0.0118} & \cellcolor{GreedyShade}\WithGreedyRatio{\textbf{0.588}}{0.0166} & \cellcolor{GreedyShade}\WithGreedyRatio{\textbf{0.229}}{0.0206} \\
 & Greedy ($\alpha=1$) & \WithGreedyRatio{0.255}{0.0732} & \WithGreedyRatio{\Second{0.188}}{0.0461} & \WithGreedyRatio{\Second{0.263}}{0.0603} & \WithGreedyRatio{\Second{0.446}}{0.0181} & \WithGreedyRatio{\Second{0.588}}{0.0268} & \WithGreedyRatio{\Second{0.225}}{0.0301} \\
\bottomrule
\end{tabular}
\par\vspace{3pt}
\raggedright\fontsize{7.8}{9.6}\selectfont
Random selects 20 queries from the 200-query reference pool; Greedy selects 20 of 200 queries ($K=20$ neighbors/query; Industrial reports partition averages).
Parentheses show mean pairwise Jaccard overlap (\textcolor{RandomRatio}{green: Random}; \textcolor{GreedyRatio}{red: Greedy}). \textbf{Bold} and \Second{dashed underlines} indicate the best and second-best Recall among the four strategies per backbone and dataset; CACO training uses $\alpha=0.5$ (shaded).
\par
\endgroup
\end{table*}

\subsection{Generalization Across Generative Backbones}
\label{sec:cross-backbone}
To test whether CACO serves as a general post-training module beyond Base, we attach it to five additional frozen continuous generative retrievers---flow-matching X-Rec~\mbox{\cite{shen2026xrec}} and diffusion-based DreamRec~\mbox{\cite{yang2023dreamrec}}, DimeRec~\mbox{\cite{li2025dimerec}}, PreferDiff~\mbox{\cite{liu2025preferdiff}}, and TA-Rec~\mbox{\cite{mao2025tarec}}---on Beauty, Sports, and Toys under the $20\times20$ budget.
Each pretrained backbone and the item embeddings remain frozen; the adjustment network takes the backbone's own history features and trains for two epochs with $M=200$ reference queries refreshed after the first epoch.

Figure~\ref{fig:cross-backbone-caco} compares paired $\operatorname{Recall}@400$ before and after CACO adjustment from identical initial queries.
CACO improves mean $\operatorname{Recall}@400$ for all six backbones across all three datasets (all 18 settings), delivering gains of 0.89--8.72 percentage points.
DimeRec achieves the largest gain on Toys, jumping from 10.38\% to 19.10\% (+8.72\,pp), while TA-Rec gains 4.49--5.12\,pp across datasets.
Even on the strongest baseline X-Rec, CACO raises $\operatorname{Recall}@400$ from 20.57\% to 23.36\% on Beauty and from 20.66\% to 24.55\% on Toys, confirming plug-and-play compatibility across flow-matching and diffusion architectures.

\subsection{Overlap Reduction and Inference Efficiency}
\label{sec:overlap-analysis}
\setcounter{topnumber}{1}
\begin{figure}[t]
  \begin{minipage}[t]{\columnwidth}
  \vspace{0pt}
  \centering
  \makebox[\linewidth][l]{\includegraphics[trim=5.75bp 4.43bp 8.12bp 8bp,clip,width=1.02\linewidth]{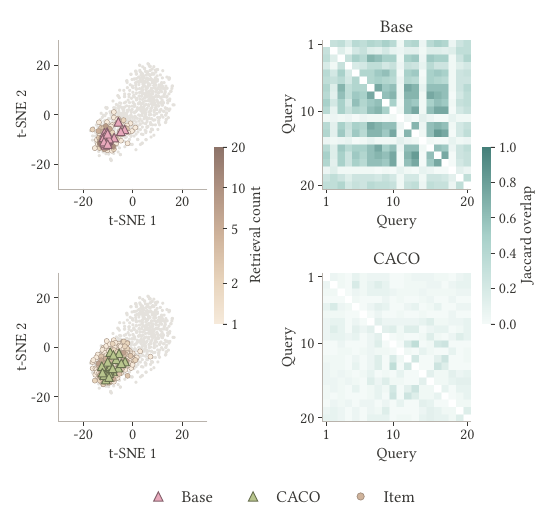}}
  \caption{An example case on Sports. Base (top) and CACO (bottom) share the same t-SNE projection of queries and items (left). Right: pairwise Jaccard overlap between the retrieval sets of queries $q_i$ and $q_j$, $J_{ij}=\frac{|\mathcal A_K(q_i)\cap\mathcal A_K(q_j)|}{|\mathcal A_K(q_j)\cup\mathcal A_K(q_i)|}$, with self-overlap masked white. Full-test averages appear in Table~\ref{tab:jaccard-overlap}.}
  \Description{One Sports case with Base on the top row and CACO on the bottom row. Left: shared t-SNE projections; rose and pistachio triangles denote Base and CACO queries, respectively. Circles denote items, with retrieved items colored by retrieval count on a shared logarithmic scale from one to twenty. Right: query retrieval-set Jaccard matrices on a shared zero-to-one scale, with diagonals masked white.}
  \label{fig:case-study}
  \end{minipage}
\end{figure}
\begin{figure}[tbp]
  \begin{minipage}[t]{\columnwidth}
    \vspace{0pt}
    \centering
    \includegraphics[width=0.98\linewidth]{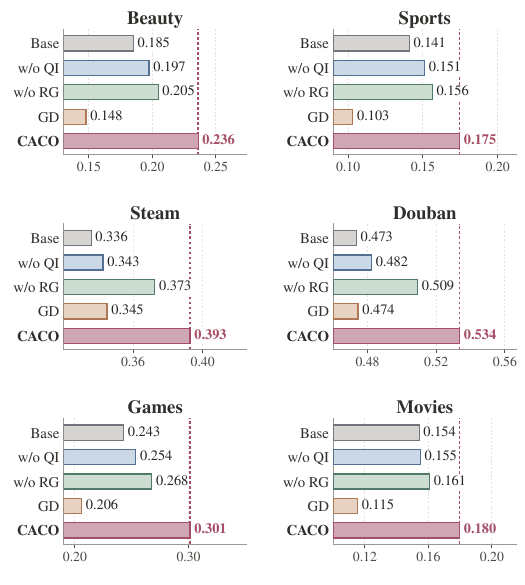}
    \caption{Ablation of CACO at Recall@400. Dashed lines mark full CACO. Horizontal axes are truncated.}
    \label{fig:ablation}
  \end{minipage}
  \Description{Six panels compare Base, removal of query interaction, removal of the relative directions gate, greedy distillation, and full CACO. Full CACO has the highest recall in each panel.}
\end{figure}
Table~\ref{tab:jaccard-overlap} reports test-set mean pairwise Jaccard overlap ($\operatorname{MPJ}$) and end-to-end inference time (query generation plus ANN retrieval at batch size 32 on an NVIDIA H100 GPU) under the $20\times20$ budget, using identical retrieval indices for Base and CACO.
CACO cuts test-set $\operatorname{MPJ}$ by 53.3\%--64.9\% across Beauty, Sports, Toys, and Industrial while adding only 0.51\%--2.97\% end-to-end latency (0.03--0.13\,ms/sample).
Grouping test users into validation-tertile Base $\operatorname{MPJ}$ strata on Beauty, Sports, and Toys further shows positive recall gains in every stratum, peaking at +6.86 to +13.23 percentage points in the high-overlap group where uncoordinated queries waste the most budget.
Moreover, expanding retrieval depth $K\in[20,400]$ on the industrial dataset under a 400-item candidate cap confirms that post-hoc deduplication cannot replace coordination: CACO reaches 20\% recall at $K=32$ (0.58\,ms CPU ANN time), whereas Base requires $K=200$ (1.37\,ms) to match it.
Across $(L,K)\in\{(10,40),\allowbreak(20,20),\allowbreak(40,10)\}$ at fixed $B=400$ on Beauty, Sports, and Toys without retraining, CACO improves $\operatorname{Recall}@400$ in all nine settings by 2.87--6.40\,pp, peaking at $40\times10$, and raises distinct items by 35.7\%--66.2\% (Table~\ref{tab:appendix-budget-transfer}). It also tops the 14-method recall--latency frontier (Figure~\ref{fig:appendix-recall-cost}).

Figure~\ref{fig:case-study} visualizes a Sports test instance via a shared t-SNE~\mbox{\cite{vandermaaten2008tsne}} projection (left) and pairwise Jaccard matrix (right).
Under Base, sampled queries cluster tightly and over 10 queries retrieve identical items ($\operatorname{MPJ}=0.306$, 104 distinct items), missing the target.
CACO spreads queries across complementary neighborhoods, lowering $\operatorname{MPJ}$ to 0.048, expanding coverage to 232 distinct items, and recovering the target item.

\subsection{Ablation Study}
\label{sec:ablation-results}
Figure~\ref{fig:ablation} compares CACO against the paired Base model and three ablated variants on Beauty, Sports, Steam, Douban, Games, and Movies under the $20\times20$ budget.
All six datasets follow $\text{Base}<\text{w/o QI}<\text{w/o RG}<\text{CACO}$.

\noindent\textbf{Effect of Query Interaction (w/o QI).}
Masking off-diagonal self-attention also removes peer relative directions in Eq.~\eqref{eq:relative-gates}. This lowers recall by 9.62\%--16.38\% relative to CACO, with the largest relative drop on Beauty (0.236 to 0.197).

\noindent\textbf{Effect of Relative Directions Gate (w/o RG).}
Removing the relative directions gate in Eq.~\eqref{eq:relative-residual} while retaining attention lowers recall by 4.59\%--13.35\%. The consistent advantage of \textbf{w/o RG} over \textbf{w/o QI} isolates the contribution from self-attention among queries.

\noindent\textbf{Marginal Coverage Learning vs.\ Greedy Distillation (GD).}
GD selects 20 teacher queries from the 200-query reference set via offline greedy selection ($\alpha=0.5$), pairs predicted and teacher queries via Hungarian matching, and minimizes their cosine distance.
GD trails CACO by 11.08\%--40.93\% in relative recall. It slightly exceeds Base on Steam and Douban but falls below Base on the four Amazon datasets, where matching teacher embeddings fails to preserve the teachers' retrieval gains.

\noindent\textbf{Effect of Reference-Pool Refresh.}
On Beauty, Sports, and Toys, refreshing $Q_h$ after epoch 1 improves mean $\operatorname{Recall}@400$ by 0.35--0.89\,pp over a two-epoch fixed-pool control, with gains for every training seed (Tables~\ref{tab:appendix-refresh}--\ref{tab:appendix-refresh-seeds}).
This isolates updated supervision gains from extra optimization, whereas additional refresh rounds plateau (Appendix~\ref{app:iterative-refinement}).

\subsection{Coverage Objective Analysis}
\label{sec:greedy-results}
Table~\ref{tab:public-greedy-table} evaluates offline greedy selection under $F_h(Q)$ across five retrieval backbones and six datasets, comparing random selection against unweighted ($\alpha=0$), sublinear ($\alpha=0.5$), and linear ($\alpha=1$) frequency-weighted coverage.
At $\alpha=0.5$ and $\alpha=1$, greedy selection improves $\operatorname{Recall}@400$ over random sampling across all 30 settings and generally lowers $\operatorname{MPJ}$. At $\alpha=0.5$, Base gains 3.25--9.33\,pp on public sets and rises from 0.181 to 0.222 on Industrial. Across Amazon, single-pass CACO (Table~\ref{tab:overall-retrieval}) surpasses Base greedy selection at $\alpha=0.5$ (0.236 vs.\ 0.216 on Beauty; 0.244 vs.\ 0.227 on Toys), as continuous tangent-space updates and pool refresh reach complementary neighborhoods beyond discrete $Q_h$ samples. Intermediate exponents $\alpha\in\{0.25,0.75\}$ also improve recall in all 30 settings (Appendix~\ref{sec:appendix-alpha-sensitivity}).

Comparing $\alpha\in\{0,0.5,1\}$ exposes \emph{relevance drift} when broadening coverage without item-value feedback.
On X-Rec, $\alpha=0$ cuts $\operatorname{MPJ}$ to 0.0000 on Steam and Douban yet drops recall (0.316 to 0.225; 0.449 to 0.335), whereas $\alpha=1$ reaches 0.413 and 0.547.
DimeRec favors smaller exponents on Amazon, where its random-pool $\operatorname{MPJ}$ is 0.622--0.668. On Steam and Douban, its initial overlap is much lower (0.057 and 0.076), and recall peaks at $\alpha=1$.
For Base and CACO, $\alpha=0.5$ achieves the highest recall on all six datasets in Table~\ref{tab:public-greedy-table}; in the finer sweep of Table~\ref{tab:appendix-greedy-alpha}, only CACO on Douban edges higher at $\alpha=0.75$ with 0.590 versus 0.588.
Finally, varying $M\in[20,800]$ on Beauty, Sports, and Toys saturates rapidly, with $M=20$ retaining 89\%--91\% of the $M=800$ gain.

\section{Conclusion}
\label{sec:conclusion}
We proposed CACO, a lightweight self-supervised post-training framework that coordinates queries from frozen flow-matching and diffusion retrievers toward complementary coverage under a fixed retrieval budget.
Using empirical reference-query retrieval counts as a relevance proxy, CACO learns coordinated query adjustment from unbiased marginal-coverage gradients without constructing a large query pool at inference.
Experiments across eight public benchmarks, an in-house streaming industrial dataset, and six generative backbones show that CACO effectively improves recall while substantially cutting inter-query overlap with negligible latency overhead.
\setcounter{topnumber}{2}

\bibliographystyle{ACM-Reference-Format}
\bibliography{references}

\clearpage
\appendix
\suppressfloats[t]
\section{Iterative Reference-Pool Updates}
\label{app:iterative-refinement}
We examine whether repeated reference-pool updates yield further gains.
Starting from Base's $M=200$ reference queries per context, we update the pool using the preceding adjustment network.
Each stage trains a new network for two epochs from the same initialization within each seed, holding its pool fixed.
Base and the initial queries $Q^{(0)}$ remain unchanged; inference uses only the current adjustment network at \mbox{$20\times20$}.

\begin{figure}[t]
\centering
\includegraphics[width=0.9\columnwidth]{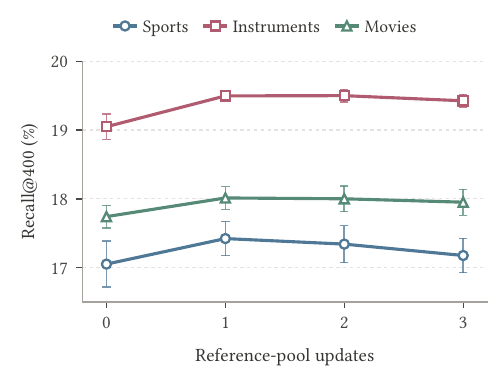}
\caption{Recall with successive reference-pool updates, starting from Base's pool. Each point averages three training seeds and two inference seeds; error bars show the standard deviation across training-seed means.}
\Description{Recall at 400 versus zero to three reference-pool updates on Sports, Instruments, and Movies. All three datasets improve after the first update, with little additional gain from later updates. Open circles, squares, and triangles distinguish the datasets; error bars show variation across training seeds.}
\label{fig:appendix-refinement-stages}
\end{figure}

Figure~\ref{fig:appendix-refinement-stages} shows that the first update improves Recall@400 by 0.37, 0.45, and 0.27 percentage points on Sports, Instruments, and Movies, respectively.
Further updates yield diminishing returns: Sports declines, while Instruments and Movies nearly plateau before decreasing slightly, despite additional rounds of training and reference generation.

\section{Reference-Query Pool Size}
\label{app:reference-pool-size}
We vary the reference-query pool size $M$ from 20 to 800 on Sports, Toys, and Beauty.
Each setting uses the same $M$ in both training epochs: we train for one epoch, refresh the pool with the resulting adjustment network, and continue training that network for one more epoch.
All other training settings remain fixed, and inference uses one adjustment network at \mbox{$20\times20$}.

\begin{figure*}[t]
\centering
\includegraphics[width=\textwidth]{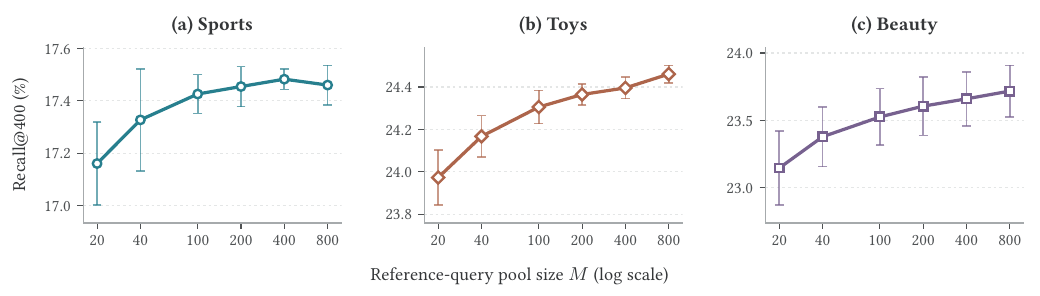}
\caption{Recall@400 versus the reference-query pool size used in both training epochs. Points average three training seeds, each evaluated with two inference seeds; error bars show the standard deviation across training-seed means. Vertical scales differ across panels.}
\Description{Three panels show Recall at 400 for reference-query pool sizes 20, 40, 100, 200, 400, and 800 on a logarithmic horizontal axis. Sports uses teal circles, Toys terracotta diamonds, and Beauty muted purple squares. Mean recall increases with diminishing returns, with a slight decline from 400 to 800 on Sports. Error bars show one standard deviation across three training-seed means. Each panel has its own nonzero vertical scale.}
\label{fig:appendix-reference-pool-size}
\end{figure*}

Figure~\ref{fig:appendix-reference-pool-size} shows diminishing returns as $M$ increases.
Sports nearly plateaus beyond $M=100$, while Toys and Beauty continue to improve modestly.
Even $M=20$ retains 89--91\% of the Recall@400 gain over Base achieved with $M=800$ across the three datasets.
One possible explanation is that the adjustment network learns from many user contexts within each dataset, reducing its reliance on large reference pools for individual contexts.

\section{Theoretical Analysis}
\label{sec:theoretical-analysis}
\setcounter{equation}{0}
\renewcommand{\theequation}{\thesection.\arabic{equation}}
We first express the marginal coverage signal as a difference in probability mass under an item distribution estimated from reference queries.
We then show that the negative gradient of the CACO training loss is an unbiased estimator of the gradient of the corresponding Gaussian-smoothed coverage objective.

\subsection{Probability-Mass Interpretation of Coverage Gains}
\label{sec:app-probability-mass}
The base generator induces an item-level retrieval probability
\begin{equation}
\begin{aligned}
r_h(v)&=\Pr_{q\sim p_{\theta_0}(\cdot\mid h)}
\bigl(v\in\mathcal A_K(q)\bigr),\\
\widehat r_h(v)&=\frac{c_h(v)}{M},
\qquad \mathbb E_{Q_h}[\widehat r_h(v)\mid h]=r_h(v).
\end{aligned}
\label{eq:app-retrieval-probability}
\end{equation}
The equality follows by taking the expectation of each retrieval indicator in Eq.~\eqref{eq:item-frequency}.
Thus, normalized retrieval counts are unbiased estimates of these probabilities.

For $\alpha>0$ and $Z_h=\sum_{v\in\mathcal I}w_h(v)>0$, the frequency weights define the empirical item distribution
\begin{equation}
\widehat p_{h,\alpha}(v)
=\frac{\widehat r_h(v)^\alpha}{\sum_{u\in\mathcal I}\widehat r_h(u)^\alpha}
=\frac{w_h(v)}{Z_h}.
\label{eq:app-empirical-item-distribution}
\end{equation}
For $\alpha=0$, this distribution is uniform over items retrieved by the reference pool.
The common factor $M^{-\alpha}$ cancels, so replacing counts by frequencies preserves the normalized coverage objective and training signal.
In particular,
\begin{equation}
\frac{F_h(Q)}{Z_h}
=\sum_{v\in\mathcal U_K(Q)}\widehat p_{h,\alpha}(v).
\label{eq:app-coverage-mass}
\end{equation}
Let $U_{-i}=\bigcup_{j\ne i}\mathcal A_K(q_j^+)$ be the items covered by the other queries.
Their contribution cancels in Eq.~\eqref{eq:counterfactual-gain}, giving
\begin{equation}
\Delta_i
=\sum_{v\in\mathcal A_K(q_i^+)\setminus U_{-i}}\widehat p_{h,\alpha}(v)
-\sum_{v\in\mathcal A_K(q_i^-)\setminus U_{-i}}\widehat p_{h,\alpha}(v).
\label{eq:app-marginal-mass-difference}
\end{equation}
Hence, $\Delta_i$ compares the probability mass newly covered by the positive and negative perturbations under the estimated item distribution.

\begin{samepage}
\begin{proposition}[Unbiased coverage difference with linear weights]
\label{prop:unbiased-coverage-mass}
For $\alpha=1$ and retrieval sets of size $K$, define $p_h^{\mathrm{ret}}(v)=r_h(v)/K$.
For any fixed pair $Q^+,Q^{(i)}$ evaluated with an independently sampled reference pool,
\begin{equation}
\begin{aligned}
\mathbb E_{Q_h}[\Delta_i\mid h,Q^+,Q^{(i)}]
&=\sum_{v\in\mathcal U_K(Q^+)}p_h^{\mathrm{ret}}(v)\\
&\quad-\sum_{v\in\mathcal U_K(Q^{(i)})}p_h^{\mathrm{ret}}(v).
\end{aligned}
\label{eq:app-unbiased-mass-difference}
\end{equation}
\end{proposition}
\end{samepage}
\begin{proof}
Each reference query retrieves $K$ items, so $\sum_v r_h(v)=K$ and $Z_h=MK$.
Consequently, $\widehat p_{h,1}(v)=c_h(v)/(MK)$ has expectation $p_h^{\mathrm{ret}}(v)$ by Eq.~\eqref{eq:app-retrieval-probability}.
Applying linearity of expectation to the difference of the two covered masses proves the result.
\end{proof}

For a refreshed pool, the same frequency construction uses the current reference-query distribution.
The following analysis establishes the training-gradient result conditional on the resulting item weights, for every supported value of $\alpha$.

\subsection{Gaussian-Smoothed Coverage Objective}
\label{sec:app-smoothed-objective}
Let $\mathcal C$ denote the fixed context $h$, frozen base outputs $(Q^{(0)},H_h)$, reference set $Q_h$, nonnegative item weights $w_h$, and a measurable retrieval rule, including its index, filtering, and tie-breaking rules.
We condition on $\mathcal C$ throughout the derivation.
Assume $0<Z_h=\sum_{v\in\mathcal I}w_h(v)<\infty$ and $\sigma>0$.
The perturbations are independent of $\mathcal C$ and independent across queries.

For each unit initial query $\dot q_i$, let
$\mathcal T_i=\{u\in\mathbb R^d:\langle u,\dot q_i\rangle=0\}$
and $\mathcal T=\mathcal T_1\times\cdots\times\mathcal T_L$.
Write $P_i=I_d-\dot q_i\dot q_i^\top$ for the matrix of the projection defined in Section~\ref{sec:query-adjustment}, and let $P$ apply these projections blockwise.
For $u=(u_1,\ldots,u_L)\in\mathcal T$, define
\begin{equation}
\begin{aligned}
\mathcal R_i(u_i)&=\frac{\dot q_i+u_i}{\sqrt{1+\|u_i\|^2}},\\
f_{\mathcal C}(u)&=
\frac{F_h\bigl((\mathcal R_i(u_i))_{i=1}^L\bigr)}{Z_h}.
\end{aligned}
\label{eq:app-normalized-coverage}
\end{equation}
Because coverage counts each item at most once, $0\leq f_{\mathcal C}(u)\leq1$.
Gaussian smoothing gives the objective
\begin{equation}
J_\sigma(\mu\mid\mathcal C)=\mathbb E_\epsilon[
f_{\mathcal C}(\mu+\sigma P\epsilon)\mid\mathcal C],
\quad \mu\in\mathcal T.
\label{eq:app-gaussian-objective}
\end{equation}
As in Eq.~\eqref{eq:perturbed-queries}, smoothing acts on tangent coordinates before query normalization, with Gaussian noise added after bounding the deterministic update in Eq.~\eqref{eq:bounded-update}.
Throughout this appendix, $\nabla_{\mu_i}J_\sigma$ denotes the gradient within $\mathcal T_i$, represented as a vector in $\mathbb R^d$.

\subsection{Unbiasedness of the Antithetic Gradient Estimator}
\label{sec:app-antithetic-proof}
\begin{theorem}[Conditional unbiasedness]
\label{thm:unbiased-coverage-gradient}
For $\epsilon_i\sim\mathcal N(0,I_d)$ independent across queries and of $\mathcal C$, and $\Delta_i$ from Eq.~\eqref{eq:counterfactual-gain},
\[
\widehat g_i=\frac{\Delta_i}{2\sigma}P_i(\epsilon_i),
\qquad
\mathbb E[\widehat g_i\mid\mathcal C]
=\nabla_{\mu_i}J_\sigma(\mu\mid\mathcal C),
\]
where the gradient is taken in the fixed tangent space at $\dot q_i$.
Moreover, for differentiable adjustment-network outputs $\mu(\phi)$, treating $\Delta_i$ as constant during backpropagation gives
\[
\mathbb E[-\nabla_\phi\mathcal L_{\mathrm{train}}\mid\mathcal C]
=\nabla_\phi J_\sigma(\mu(\phi)\mid\mathcal C).
\]
\end{theorem}

The proof uses Gaussian smoothing in tangent coordinates and symmetry of the paired perturbations; discrete retrieval need only be measurable.

Choose an orthonormal basis $B_i\in\mathbb R^{d\times(d-1)}$ for $\mathcal T_i$, so that $B_iB_i^\top=P_i$ and $B_i^\top B_i=I_{d-1}$.
Let $B=\operatorname{diag}(B_1,\ldots,B_L)$, $\mu=Bx$, and $\xi=B^\top\epsilon\sim\mathcal N(0,I_m)$, where $m=L(d-1)$.
Then $B\xi=P\epsilon$ and Eq.~\eqref{eq:app-gaussian-objective} is an ordinary Gaussian convolution on $\mathbb R^m$.
Writing $\varphi_{\sigma,m}$ for the density of $\mathcal N(0,\sigma^2I_m)$ yields
\begin{equation}
\begin{aligned}
\overline J_\sigma(x)
&=\int_{\mathbb R^m} f_{\mathcal C}(Bz)
\varphi_{\sigma,m}(z-x)\,\mathrm dz,\\
\nabla_{x_i}\overline J_\sigma(x)
&=\frac{1}{\sigma}\mathbb E_\xi[
f_{\mathcal C}(Bx+\sigma B\xi)\xi_i\mid\mathcal C].
\end{aligned}
\label{eq:app-gaussian-score}
\end{equation}
Here $\overline J_\sigma(x)=J_\sigma(Bx\mid\mathcal C)$, and the gradient identity follows from
$\nabla_{x_i}\varphi_{\sigma,m}(z-x)=(z_i-x_i)\varphi_{\sigma,m}(z-x)/\sigma^2$
and the substitution $z=x+\sigma\xi$.
The boundedness of $f_{\mathcal C}$ and integrability of Gaussian density derivatives justify differentiation under the integral.
This Gaussian score identity underlies evolution strategies~\cite{salimans2017es} and applies even when the retrieval sets change discontinuously.
Multiplying the second line by $B_i$ gives
\begin{equation}
\nabla_{\mu_i}J_\sigma(\mu\mid\mathcal C)
=\frac{1}{\sigma Z_h}
\mathbb E_\epsilon[F_h(Q^+)P_i\epsilon_i\mid\mathcal C].
\label{eq:app-tangent-score}
\end{equation}

Next, let $\epsilon^{(i)}$ equal $\epsilon$ with only block $i$ negated.
Independence and Gaussian symmetry imply that $\epsilon^{(i)}$ and $\epsilon$ have the same conditional distribution.
Moreover, the single-query replacement in Eq.~\eqref{eq:counterfactual-gain} satisfies
$Q^{(i)}(\mu,\epsilon)=Q^+(\mu,\epsilon^{(i)})$.
Changing variables from $\epsilon$ to $\epsilon^{(i)}$ therefore gives
\begin{equation}
\mathbb E[F_h(Q^{(i)})P_i\epsilon_i\mid\mathcal C]
=-\mathbb E[F_h(Q^+)P_i\epsilon_i\mid\mathcal C].
\label{eq:app-antithetic-reflection}
\end{equation}
Combining this equality with Eqs.~\eqref{eq:counterfactual-gain} and~\eqref{eq:app-tangent-score} proves
\begin{equation}
\begin{aligned}
\mathbb E[\widehat g_i\mid\mathcal C]
&=\frac{1}{2\sigma Z_h}\mathbb E\!\left[
\bigl(F_h(Q^+)-F_h(Q^{(i)})\bigr)P_i\epsilon_i
\,\middle|\,\mathcal C\right]\\
&=\nabla_{\mu_i}J_\sigma(\mu\mid\mathcal C).
\end{aligned}
\label{eq:app-query-unbiasedness}
\end{equation}
The other queries retain their positive perturbations in both collections and are averaged over in $J_\sigma$.
Thus each estimate corresponds to a partial gradient of the jointly smoothed objective; sharing $Q^+$ may correlate the estimates but preserves their unbiasedness.

\subsection{CACO Parameter Gradient}
\label{sec:app-parameter-gradient}
Let $\phi$ denote the adjustment-network parameters and $D_\phi\mu_i$ its Jacobian at a differentiable point.
Conditional on $\mathcal C$, the initial queries and projections $P_i$ are independent of $\phi$.
For the bounded update in Eq.~\eqref{eq:bounded-update}, set
$a_i=(1+\|t_i\|^2/\rho^2)^{-1/2}$.
Since $t_i=P_i r_i(\phi)$ and $\mu_i=a_i t_i$, the chain rule gives
\begin{equation}
D_\phi\mu_i=
\left(a_i I_d-\frac{a_i^3}{\rho^2}t_it_i^\top\right)
P_iD_\phi r_i.
\label{eq:app-bounded-jacobian}
\end{equation}
Its columns lie in $\mathcal T_i$, so
$(D_\phi\mu_i)^\top P_i=(D_\phi\mu_i)^\top$.
Treating the sampled noise and coverage differences as constants during backpropagation, as in Eq.~\eqref{eq:coverage-loss}, yields
\begin{equation}
\begin{aligned}
-\nabla_\phi\mathcal L_{\mathrm{train}}
&=\sum_{i=1}^L(D_\phi\mu_i)^\top
\frac{\Delta_i\epsilon_i}{2\sigma}\\
&=\sum_{i=1}^L(D_\phi\mu_i)^\top\widehat g_i.
\end{aligned}
\label{eq:app-surrogate-gradient}
\end{equation}
The Jacobian removes the radial component of $\epsilon_i$, which explains why Eq.~\eqref{eq:coverage-loss} can use the unprojected noise.
Since network outputs are computed before sampling perturbations, their Jacobians can be taken outside the conditional expectation.
Using Eq.~\eqref{eq:app-query-unbiasedness} and the chain rule once more gives
\begin{equation}
\begin{aligned}
\mathbb E[-\nabla_\phi\mathcal L_{\mathrm{train}}\mid\mathcal C]
&=\sum_{i=1}^L(D_\phi\mu_i)^\top
\nabla_{\mu_i}J_\sigma(\mu\mid\mathcal C)\\
&=\nabla_\phi J_\sigma(\mu(\phi)\mid\mathcal C),
\end{aligned}
\label{eq:app-parameter-unbiasedness}
\end{equation}
which completes the proof of Theorem~\ref{thm:unbiased-coverage-gradient}.
Inter-query dependencies are captured by $D_\phi\mu_i$; the network outputs need not be independent.

\paragraph{Effect of reference refresh.}
At training step $t$, let $\mathcal C_t$ contain the current reference set and weights, including any refresh completed before the new perturbations are sampled.
The reference queries and item weights are fixed during differentiation, and the new perturbations are sampled independently of $\mathcal C_t$.
Theorem~\ref{thm:unbiased-coverage-gradient} therefore gives an unbiased gradient estimate for the Gaussian-smoothed coverage objective defined by the current empirical item distribution, before and after reference refresh.

\section{Offline Greedy Selection and Sensitivity to \texorpdfstring{$\alpha$}{alpha}}
\label{sec:appendix-greedy}

\subsection{Greedy Selection Procedure}
\label{sec:appendix-greedy-alg}
Algorithm~\ref{alg:offline-greedy-coverage} selects $L$ queries from a reference pool of size $M$ ($1\leq L\leq M$) to maximize the frequency-weighted coverage objective $F_h(Q)$ defined in Section~\ref{sec:coverage-objective}.
With candidate sets $C_j=\mathcal A_K(\tilde q_j)$ and item weights $w_h$ fixed, each iteration adds the remaining query with the largest marginal weight of newly covered items and updates the covered union $U$.
The procedure requires no ground-truth target labels and preserves retrieval depth $K$; its inference cost stems from generating all $M$ reference queries and executing $M$ nearest-neighbor searches, motivating CACO's amortized single-pass adjustment of $L$ queries.

\begin{algorithm}[!ht]
\caption{Offline Greedy Query Selection}
\label{alg:offline-greedy-coverage}
\begin{algorithmic}[1]
\Require Reference queries $\tilde q_1,\ldots,\tilde q_M$; retrieval sets
         $C_j=\mathcal A_K(\tilde q_j)$;
\Statex \hspace{\algorithmicindent}item weights $w_h$; number of selected queries $L$.
\Ensure Selected query collection $Q$.
\State $Q\gets[\,]$;
\State $U\gets\varnothing$; \Comment{Covered items}
\State $R\gets\{1,\ldots,M\}$; \Comment{Unselected query indices}
\For{$t=1,\ldots,L$}
    \State \textbf{For each} $j\in R$, compute:
    \State \hspace{\algorithmicindent}$\operatorname{gain}(j)\gets\displaystyle\sum_{v\in C_j\setminus U}w_h(v)$;
    \State $j^*\gets\operatorname*{arg\,max}_{j\in R}\operatorname{gain}(j)$;
    \State $Q\gets\operatorname{Append}(Q,\tilde q_{j^*})$;
    \State $U\gets U\cup C_{j^*}$;
    \State $R\gets R\setminus\{j^*\}$;
\EndFor
\State \Return $Q$;
\end{algorithmic}
\end{algorithm}

\begin{table*}[tp]
\caption{Full sensitivity of offline greedy query selection to the relevance-weighting exponent $\alpha$: Recall@400 ($\uparrow$) and $\operatorname{MPJ}$ ($\downarrow$, parentheses; Eq.~\eqref{eq:jaccard-overlap}).}
\label{tab:appendix-greedy-alpha}
\definecolor{GreedyShade}{HTML}{FBEEDE}
\definecolor{RandomRatio}{HTML}{276347}
\definecolor{GreedyRatio}{HTML}{A12C43}
\providecommand{\Second}[1]{\begingroup\setbox0=\hbox{#1}\leavevmode\rlap{\raisebox{-1.4pt}[0pt][0pt]{\hdashrule{\wd0}{0.35pt}{1.4pt 1pt}}}\box0\endgroup}
\providecommand{\DatasetHeader}[1]{\makebox[22pt][c]{\textbf{#1}}\hspace{1.4pt}\makebox[34pt][r]{}}
\providecommand{\WithRandomRatio}[2]{\makebox[22pt][c]{#1}\hspace{1.4pt}\makebox[34pt][r]{\textcolor{RandomRatio}{\fontsize{7.5}{9}\selectfont\bfseries (#2)}}}
\providecommand{\WithGreedyRatio}[2]{\makebox[22pt][c]{#1}\hspace{1.4pt}\makebox[34pt][r]{\textcolor{GreedyRatio}{\fontsize{7.5}{9}\selectfont\bfseries (#2)}}}
\begingroup
\centering
\fontsize{9}{10.5}\selectfont
\renewcommand{\arraystretch}{1.03}
\setlength{\tabcolsep}{0pt}
\begin{tabular}{@{}p{18mm}p{27mm}*{6}{>{\raggedleft\arraybackslash}p{\dimexpr(\linewidth-45mm)/6\relax}}@{}}
\toprule
\textbf{Backbone} & \textbf{Selection} & \DatasetHeader{Beauty} & \DatasetHeader{Sports} & \DatasetHeader{Toys} & \DatasetHeader{Steam} & \DatasetHeader{Douban} & \DatasetHeader{Industrial} \\
\midrule
\multirow{6}{*}{LATTE} & Random & \WithRandomRatio{0.176}{0.0332} & \WithRandomRatio{0.110}{0.0192} & \WithRandomRatio{0.131}{0.0115} & \WithRandomRatio{0.350}{0.0375} & \WithRandomRatio{0.480}{0.0545} & \WithRandomRatio{0.098}{0.0333} \\
 & Greedy ($\alpha=0$) & \WithGreedyRatio{0.118}{0.0001} & \WithGreedyRatio{0.072}{0.0000} & \WithGreedyRatio{0.091}{0.0000} & \WithGreedyRatio{0.392}{0.0038} & \WithGreedyRatio{0.507}{0.0038} & \WithGreedyRatio{0.091}{0.0064} \\
 & Greedy ($\alpha=0.25$) & \WithGreedyRatio{0.177}{0.0012} & \WithGreedyRatio{0.117}{0.0013} & \WithGreedyRatio{0.146}{0.0010} & \WithGreedyRatio{0.410}{0.0059} & \WithGreedyRatio{0.512}{0.0056} & \WithGreedyRatio{0.139}{0.0068} \\
 & \cellcolor{GreedyShade}Greedy ($\alpha=0.5$) & \cellcolor{GreedyShade}\WithGreedyRatio{0.200}{0.0042} & \cellcolor{GreedyShade}\WithGreedyRatio{0.131}{0.0042} & \cellcolor{GreedyShade}\WithGreedyRatio{0.165}{0.0033} & \cellcolor{GreedyShade}\WithGreedyRatio{0.424}{0.0094} & \cellcolor{GreedyShade}\WithGreedyRatio{0.539}{0.0092} & \cellcolor{GreedyShade}\WithGreedyRatio{0.142}{0.0075} \\
 & Greedy ($\alpha=0.75$) & \WithGreedyRatio{\Second{0.212}}{0.0082} & \WithGreedyRatio{\Second{0.137}}{0.0075} & \WithGreedyRatio{\Second{0.175}}{0.0060} & \WithGreedyRatio{\textbf{0.427}}{0.0126} & \WithGreedyRatio{\Second{0.550}}{0.0129} & \WithGreedyRatio{\Second{0.144}}{0.0082} \\
 & Greedy ($\alpha=1$) & \WithGreedyRatio{\textbf{0.218}}{0.0125} & \WithGreedyRatio{\textbf{0.142}}{0.0108} & \WithGreedyRatio{\textbf{0.179}}{0.0087} & \WithGreedyRatio{\Second{0.426}}{0.0154} & \WithGreedyRatio{\textbf{0.554}}{0.0163} & \WithGreedyRatio{\textbf{0.144}}{0.0087} \\
\noalign{\vskip3pt\hbox{\color[HTML]{666666}\hdashrule{\linewidth}{0.6pt}{4pt 2pt}}\vskip3pt}
\multirow{6}{*}{X-Rec} & Random & \WithRandomRatio{0.205}{0.1119} & \WithRandomRatio{0.127}{0.0371} & \WithRandomRatio{0.204}{0.1316} & \WithRandomRatio{0.316}{0.0193} & \WithRandomRatio{0.449}{0.0324} & \WithRandomRatio{0.197}{0.0042} \\
 & Greedy ($\alpha=0$) & \WithGreedyRatio{0.224}{0.0419} & \WithGreedyRatio{0.105}{0.0043} & \WithGreedyRatio{0.234}{0.0644} & \WithGreedyRatio{0.225}{0.0000} & \WithGreedyRatio{0.335}{0.0000} & \WithGreedyRatio{0.187}{0.0095} \\
 & Greedy ($\alpha=0.25$) & \WithGreedyRatio{0.237}{0.0452} & \WithGreedyRatio{0.139}{0.0070} & \WithGreedyRatio{\textbf{0.240}}{0.0689} & \WithGreedyRatio{0.364}{0.0017} & \WithGreedyRatio{0.499}{0.0012} & \WithGreedyRatio{0.213}{0.0105} \\
 & \cellcolor{GreedyShade}Greedy ($\alpha=0.5$) & \cellcolor{GreedyShade}\WithGreedyRatio{\textbf{0.240}}{0.0503} & \cellcolor{GreedyShade}\WithGreedyRatio{0.150}{0.0114} & \cellcolor{GreedyShade}\WithGreedyRatio{0.239}{0.0745} & \cellcolor{GreedyShade}\WithGreedyRatio{0.393}{0.0043} & \cellcolor{GreedyShade}\WithGreedyRatio{0.526}{0.0029} & \cellcolor{GreedyShade}\WithGreedyRatio{0.217}{0.0128} \\
 & Greedy ($\alpha=0.75$) & \WithGreedyRatio{\Second{0.240}}{0.0561} & \WithGreedyRatio{\Second{0.154}}{0.0157} & \WithGreedyRatio{\Second{0.239}}{0.0805} & \WithGreedyRatio{\Second{0.406}}{0.0068} & \WithGreedyRatio{\Second{0.540}}{0.0047} & \WithGreedyRatio{\Second{0.218}}{0.0159} \\
 & Greedy ($\alpha=1$) & \WithGreedyRatio{0.239}{0.0614} & \WithGreedyRatio{\textbf{0.155}}{0.0194} & \WithGreedyRatio{0.237}{0.0860} & \WithGreedyRatio{\textbf{0.413}}{0.0091} & \WithGreedyRatio{\textbf{0.547}}{0.0066} & \WithGreedyRatio{\textbf{0.218}}{0.0191} \\
\noalign{\vskip3pt\hbox{\color[HTML]{666666}\hdashrule{\linewidth}{0.6pt}{4pt 2pt}}\vskip3pt}
\multirow{6}{*}{DimeRec} & Random & \WithRandomRatio{0.061}{0.6220} & \WithRandomRatio{0.034}{0.6675} & \WithRandomRatio{0.107}{0.6535} & \WithRandomRatio{0.339}{0.0566} & \WithRandomRatio{0.351}{0.0764} & \WithRandomRatio{0.014}{0.0257} \\
 & Greedy ($\alpha=0$) & \WithGreedyRatio{\textbf{0.079}}{0.5393} & \WithGreedyRatio{\Second{0.042}}{0.5998} & \WithGreedyRatio{\textbf{0.126}}{0.5813} & \WithGreedyRatio{0.313}{0.0035} & \WithGreedyRatio{0.296}{0.0025} & \WithGreedyRatio{0.015}{0.0024} \\
 & Greedy ($\alpha=0.25$) & \WithGreedyRatio{\Second{0.078}}{0.5407} & \WithGreedyRatio{0.042}{0.5993} & \WithGreedyRatio{\Second{0.126}}{0.5810} & \WithGreedyRatio{0.374}{0.0064} & \WithGreedyRatio{0.385}{0.0051} & \WithGreedyRatio{0.017}{0.0046} \\
 & \cellcolor{GreedyShade}Greedy ($\alpha=0.5$) & \cellcolor{GreedyShade}\WithGreedyRatio{0.078}{0.5405} & \cellcolor{GreedyShade}\WithGreedyRatio{\textbf{0.042}}{0.5986} & \cellcolor{GreedyShade}\WithGreedyRatio{0.125}{0.5801} & \cellcolor{GreedyShade}\WithGreedyRatio{0.398}{0.0121} & \cellcolor{GreedyShade}\WithGreedyRatio{0.412}{0.0101} & \cellcolor{GreedyShade}\WithGreedyRatio{\textbf{0.018}}{0.0085} \\
 & Greedy ($\alpha=0.75$) & \WithGreedyRatio{0.078}{0.5411} & \WithGreedyRatio{0.042}{0.5986} & \WithGreedyRatio{0.125}{0.5801} & \WithGreedyRatio{\Second{0.406}}{0.0184} & \WithGreedyRatio{\Second{0.422}}{0.0156} & \WithGreedyRatio{0.018}{0.0125} \\
 & Greedy ($\alpha=1$) & \WithGreedyRatio{0.078}{0.5419} & \WithGreedyRatio{0.042}{0.5987} & \WithGreedyRatio{0.125}{0.5802} & \WithGreedyRatio{\textbf{0.406}}{0.0238} & \WithGreedyRatio{\textbf{0.423}}{0.0209} & \WithGreedyRatio{\Second{0.018}}{0.0161} \\
\noalign{\vskip3pt\hbox{\color[HTML]{666666}\hdashrule{\linewidth}{0.6pt}{4pt 2pt}}\vskip3pt}
\multirow{6}{*}{Base} & Random & \WithRandomRatio{0.183}{0.2373} & \WithRandomRatio{0.136}{0.1844} & \WithRandomRatio{0.192}{0.1967} & \WithRandomRatio{0.336}{0.1047} & \WithRandomRatio{0.473}{0.2303} & \WithRandomRatio{0.181}{0.1250} \\
 & Greedy ($\alpha=0$) & \WithGreedyRatio{0.214}{0.1402} & \WithGreedyRatio{0.167}{0.0616} & \WithGreedyRatio{0.225}{0.1191} & \WithGreedyRatio{0.393}{0.0118} & \WithGreedyRatio{0.544}{0.0313} & \WithGreedyRatio{0.200}{0.0339} \\
 & Greedy ($\alpha=0.25$) & \WithGreedyRatio{\Second{0.215}}{0.1441} & \WithGreedyRatio{\Second{0.169}}{0.0660} & \WithGreedyRatio{\Second{0.227}}{0.1240} & \WithGreedyRatio{0.421}{0.0151} & \WithGreedyRatio{0.559}{0.0332} & \WithGreedyRatio{0.220}{0.0355} \\
 & \cellcolor{GreedyShade}Greedy ($\alpha=0.5$) & \cellcolor{GreedyShade}\WithGreedyRatio{\textbf{0.216}}{0.1489} & \cellcolor{GreedyShade}\WithGreedyRatio{\textbf{0.170}}{0.0721} & \cellcolor{GreedyShade}\WithGreedyRatio{\textbf{0.227}}{0.1296} & \cellcolor{GreedyShade}\WithGreedyRatio{\textbf{0.429}}{0.0205} & \cellcolor{GreedyShade}\WithGreedyRatio{\textbf{0.565}}{0.0363} & \cellcolor{GreedyShade}\WithGreedyRatio{\textbf{0.222}}{0.0392} \\
 & Greedy ($\alpha=0.75$) & \WithGreedyRatio{0.215}{0.1547} & \WithGreedyRatio{0.169}{0.0794} & \WithGreedyRatio{0.226}{0.1357} & \WithGreedyRatio{\Second{0.428}}{0.0272} & \WithGreedyRatio{\Second{0.565}}{0.0407} & \WithGreedyRatio{\Second{0.222}}{0.0441} \\
 & Greedy ($\alpha=1$) & \WithGreedyRatio{0.213}{0.1601} & \WithGreedyRatio{0.167}{0.0866} & \WithGreedyRatio{0.224}{0.1414} & \WithGreedyRatio{0.424}{0.0337} & \WithGreedyRatio{0.565}{0.0457} & \WithGreedyRatio{0.221}{0.0493} \\
\noalign{\vskip3pt\hbox{\color[HTML]{666666}\hdashrule{\linewidth}{0.6pt}{4pt 2pt}}\vskip3pt}
\multirow{6}{*}{\textbf{CACO}} & Random & \WithRandomRatio{0.224}{0.1108} & \WithRandomRatio{0.165}{0.0715} & \WithRandomRatio{0.231}{0.0864} & \WithRandomRatio{0.368}{0.0387} & \WithRandomRatio{0.517}{0.0889} & \WithRandomRatio{0.194}{0.0635} \\
 & Greedy ($\alpha=0$) & \WithGreedyRatio{0.255}{0.0527} & \WithGreedyRatio{0.183}{0.0253} & \WithGreedyRatio{0.258}{0.0387} & \WithGreedyRatio{0.408}{0.0053} & \WithGreedyRatio{0.538}{0.0099} & \WithGreedyRatio{0.217}{0.0130} \\
 & Greedy ($\alpha=0.25$) & \WithGreedyRatio{\Second{0.258}}{0.0574} & \WithGreedyRatio{0.187}{0.0298} & \WithGreedyRatio{0.262}{0.0436} & \WithGreedyRatio{0.445}{0.0082} & \WithGreedyRatio{0.577}{0.0124} & \WithGreedyRatio{\Second{0.228}}{0.0159} \\
 & \cellcolor{GreedyShade}Greedy ($\alpha=0.5$) & \cellcolor{GreedyShade}\WithGreedyRatio{\textbf{0.258}}{0.0627} & \cellcolor{GreedyShade}\WithGreedyRatio{\textbf{0.190}}{0.0355} & \cellcolor{GreedyShade}\WithGreedyRatio{\textbf{0.265}}{0.0494} & \cellcolor{GreedyShade}\WithGreedyRatio{\textbf{0.451}}{0.0118} & \cellcolor{GreedyShade}\WithGreedyRatio{\Second{0.588}}{0.0166} & \cellcolor{GreedyShade}\WithGreedyRatio{\textbf{0.229}}{0.0206} \\
 & Greedy ($\alpha=0.75$) & \WithGreedyRatio{0.256}{0.0683} & \WithGreedyRatio{\Second{0.189}}{0.0412} & \WithGreedyRatio{\Second{0.264}}{0.0553} & \WithGreedyRatio{\Second{0.450}}{0.0152} & \WithGreedyRatio{\textbf{0.590}}{0.0217} & \WithGreedyRatio{0.227}{0.0255} \\
 & Greedy ($\alpha=1$) & \WithGreedyRatio{0.255}{0.0732} & \WithGreedyRatio{0.188}{0.0461} & \WithGreedyRatio{0.263}{0.0603} & \WithGreedyRatio{0.446}{0.0181} & \WithGreedyRatio{0.588}{0.0268} & \WithGreedyRatio{0.225}{0.0301} \\
\bottomrule
\end{tabular}
\par\vspace{5pt}
\raggedright\fontsize{8}{10}\selectfont
Random selects 20 queries at random from the reference query pool; Greedy selects 20 of 200 queries ($K=20$ neighbors/query; Industrial reports partition averages).\\
Parentheses: Jaccard overlap averaged over 190 query pairs, then test contexts. \textcolor{RandomRatio}{Green: baseline}; \textcolor{GreedyRatio}{red: Greedy}.\\
\textbf{Bold} and \Second{dashed underlines} indicate the best and second-best Recall per backbone and dataset. For CACO training, we set $\alpha=0.5$ (shaded).
\par
\endgroup
\end{table*}

\subsection{Sensitivity to the Relevance-Weighting Exponent \texorpdfstring{$\alpha$}{alpha}}
\label{sec:appendix-alpha-sensitivity}
Table~\ref{tab:appendix-greedy-alpha} reports the full sensitivity sweep over $\alpha\in\{0,\allowbreak 0.25,\allowbreak 0.5,\allowbreak 0.75,\allowbreak 1\}$ across five retrieval backbones and six datasets, complementing the condensed summary in Table~\ref{tab:public-greedy-table}.

\noindent\textbf{Robustness across positive exponents.}
Every tested positive value $\alpha\in\{0.25,0.5,0.75,1\}$ improves $\operatorname{Recall}@400$ over random reference-pool selection across all 30 backbone--dataset pairs. Most settings also show lower $\operatorname{MPJ}$.

\noindent\textbf{Coverage--relevance trade-off trajectory.}
Higher values of $\alpha$ give greater weight to frequently retrieved candidates and generally permit more pairwise overlap.
On Steam and Douban, unweighted coverage ($\alpha=0$) drives X-Rec's $\operatorname{MPJ}$ to 0.0000, but reduces recall from 0.316 to 0.225 and from 0.449 to 0.335, respectively.
At $\alpha=0.25$, recall already exceeds random selection on both datasets, reaching 0.364 and 0.499, respectively.

\noindent\textbf{Link between preferred $\alpha$ and intrinsic query redundancy.}
Comparing optimal exponents across backbones reveals a clear relationship with initial retrieval overlap.
Highly redundant generators such as DimeRec (random-pool $\operatorname{MPJ}$ of $0.6220$--$0.6675$ on Amazon benchmarks) peak at smaller exponents ($\alpha\in[0,0.5]$), where dispersing collapsed neighborhoods is the primary bottleneck.
For the same DimeRec backbone on Steam and Douban, initial $\operatorname{MPJ}$ is only 0.0566 and 0.0764, and $\alpha=1$ yields the best recall on both. X-Rec also peaks at $\alpha=1$ on these two datasets.
Base and CACO exhibit moderate redundancy and achieve optimal or near-optimal recall at $\alpha=0.5$ across datasets, supporting our default choice.

\begingroup
\setlength{\dblfloatsep}{8pt}
\section{Additional Qualitative Cases}
\label{sec:appendix-cases}
\begin{table*}[!tp]
\caption{Selected success cases under $20\times20$ retrieval, shown with inference seed 42. Arrows denote Base $\rightarrow$ CACO. Best target rank is the minimum rank across queries; ranks at most 20 indicate a hit.}
\label{tab:appendix-cases}
\centering
\begin{tabular}{llllll}
\toprule
Example & Dataset & Target item & Distinct items & $\operatorname{MPJ}$ & Best target rank \\
\midrule
Example 1 & Beauty & Nail-art transfer foil & $66\rightarrow201$ & $0.353\rightarrow0.077$ & $88\rightarrow3$ \\
Example 2 & Beauty & Concealer & $116\rightarrow232$ & $0.296\rightarrow0.057$ & $28\rightarrow3$ \\
Example 3 & Sports & Cycling computer & $112\rightarrow241$ & $0.243\rightarrow0.047$ & $27\rightarrow3$ \\
Example 4 & Toys & Cards Against Humanity & $161\rightarrow311$ & $0.106\rightarrow0.020$ & $49\rightarrow2$ \\
Example 5 & Toys & The Castles of Burgundy & $89\rightarrow222$ & $0.292\rightarrow0.054$ & $22\rightarrow3$ \\
\bottomrule
\end{tabular}
\end{table*}

The five cases in Table~\ref{tab:appendix-cases} illustrate how coordinated query adjustment can reduce retrieval overlap and recover relevant items missed by Base.
They are selected successes from Amazon Reviews 2014 Beauty, Sports, and Toys: CACO retrieves the target under both inference seeds 42 and 43 and returns more distinct items.
These examples complement the aggregate results and Figure~\ref{fig:case-study}.

Across all five cases, the target moves from outside every query's top-20 results to rank 2 or 3 for at least one query.
The improvement also appears in the Toys case with relatively low initial overlap, where $\operatorname{MPJ}$ decreases from $0.106$ to $0.020$.
CACO can replace some items retrieved by Base, so the broader candidate set need not contain the original set.

The examples use 8M Base and CACO models trained with seed 42, with $L=20$ queries and retrieval depth $K=20$.
Previously interacted items are filtered, and the merged candidate sets are not refilled.
Table~\ref{tab:appendix-cases} reports inference seed 42; item counts, $\operatorname{MPJ}$, and target ranks are computed in the original embedding space.

Figures~\ref{fig:app-beauty-12525}--\ref{fig:app-toys-15365} use a shared t-SNE projection within each case, containing both query collections, their retrieved-item union, recent history, the target, and 500 background items.
The projection uses cosine distance, perplexity 35, and seed 42.
Base appears above CACO, with query distributions on the left and Jaccard matrices on the right.
All panels preserve query order and share color scales: logarithmic retrieval counts from 1 to 20 and Jaccard overlap from 0 to 1.
The heatmap diagonal is masked and excluded from $\operatorname{MPJ}$; projected distances serve only to illustrate the retrieval geometry.

\begin{figure}[!ht]
\centering
\includegraphics[width=0.94\linewidth]{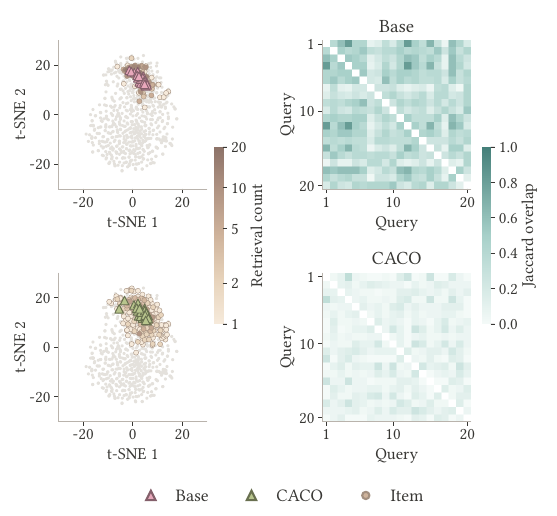}
\caption{Example 1 (Beauty): nail-art transfer foil. CACO broadens coverage and reduces query overlap, bringing the target into the retrieved set.}
\Description{Example 1, a selected Beauty success case. Base and CACO have 66 and 201 distinct items, with mean pairwise Jaccard overlap 0.353 and 0.077. Triangles denote queries and circles denote items.}
\label{fig:app-beauty-12525}
\end{figure}

\begin{figure*}[!tp]
\begin{minipage}[t]{\columnwidth}
\vspace{0pt}
\centering
\includegraphics[width=0.94\linewidth]{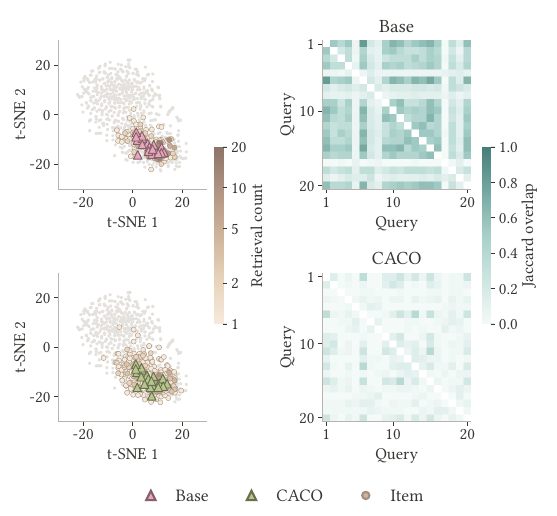}
\caption{Example 2 (Beauty): concealer. Query adjustment reduces overlap and recovers the target missed by Base.}
\Description{Example 2, a selected Beauty success case. Base and CACO have 116 and 232 distinct items, with mean pairwise Jaccard overlap 0.296 and 0.057. Distribution panels are on the left and overlap matrices on the right.}
\label{fig:app-beauty-5535}
\end{minipage}\hfill
\begin{minipage}[t]{\columnwidth}
\vspace{0pt}
\centering
\includegraphics[width=0.94\linewidth]{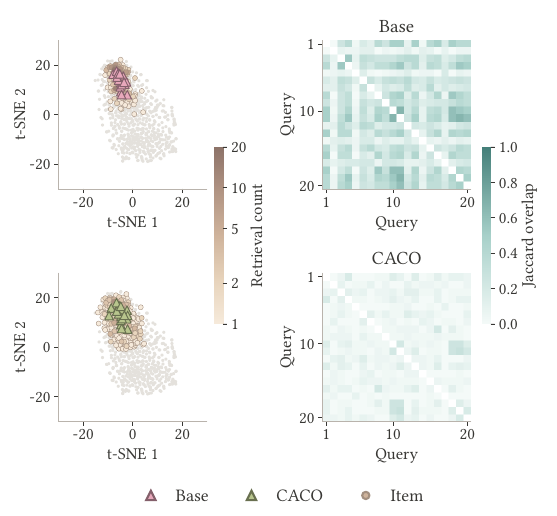}
\caption{Example 3 (Sports): cycling computer. CACO retrieves the target while covering more distinct items.}
\Description{Example 3, a selected Sports success case. Base and CACO have 112 and 241 distinct items, with mean pairwise Jaccard overlap 0.243 and 0.047.}
\label{fig:app-sports-3209}
\end{minipage}
\par\medskip
\begin{minipage}[t]{\columnwidth}
\vspace{0pt}
\centering
\includegraphics[width=0.94\linewidth]{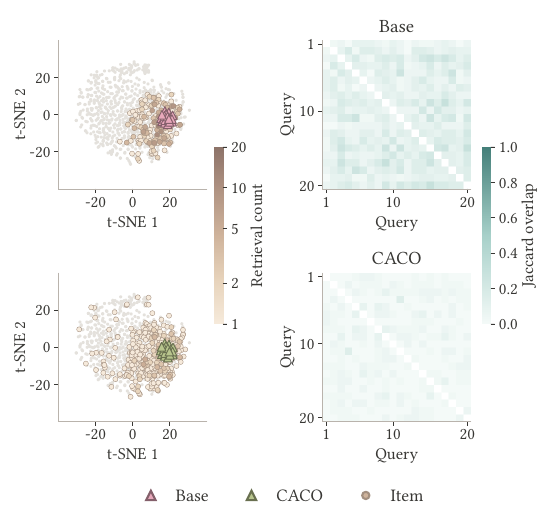}
\caption{Example 4 (Toys): Cards Against Humanity. CACO further reduces the relatively low initial overlap and retrieves the target.}
\Description{Example 4, a selected Toys success case. Base and CACO have 161 and 311 distinct items, with mean pairwise Jaccard overlap 0.106 and 0.020.}
\label{fig:app-toys-7948}
\end{minipage}\hfill
\begin{minipage}[t]{\columnwidth}
\vspace{0pt}
\centering
\includegraphics[width=0.94\linewidth]{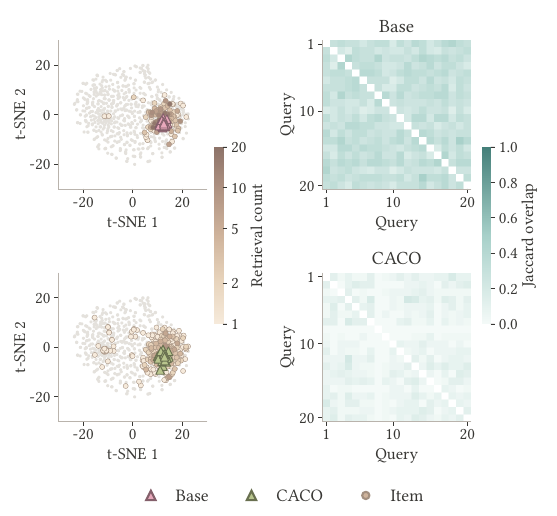}
\caption{Example 5 (Toys): The Castles of Burgundy. Reduced overlap accompanies broader coverage and successful target retrieval.}
\Description{Example 5, a selected Toys success case. Base and CACO have 89 and 222 distinct items, with mean pairwise Jaccard overlap 0.292 and 0.054.}
\label{fig:app-toys-15365}
\end{minipage}
\end{figure*}
\endgroup

\section{Additional Experiments}
\label{app:additional-experiments}
\setcounter{dbltopnumber}{2}
\renewcommand{\dbltopfraction}{0.8}
\subsection{Effect of Reference-Pool Refresh}
\label{app:refresh}
\begin{table*}[t]
\centering
\caption{Effect of reference-pool refresh on recall (\%). Values are mean $\pm$ sample standard deviation over three training seeds, each averaged over two inference seeds. $\Delta_{\mathrm{Recall}}$ is the paired gain from refreshing after the first epoch rather than keeping the pool fixed for two epochs, in percentage points. Training uses $20\times20$; $10\times10$ results are obtained without retraining.}
\label{tab:appendix-refresh}
\small
\setlength{\tabcolsep}{8pt}
\begin{tabular}{llrrrr}
\toprule
Dataset & $L\times K$ & 1 epoch & 2 epochs, fixed pool & 2 epochs, refreshed pool & $\Delta_{\mathrm{Recall}}$ (pp) \\
\midrule
Beauty & $20\times20$ & $22.498\pm0.332$ & $22.896\pm0.340$ & $23.606\pm0.217$ & $+0.710\pm0.131$ \\
Sports & $20\times20$ & $16.835\pm0.116$ & $17.100\pm0.192$ & $17.455\pm0.076$ & $+0.354\pm0.136$ \\
Toys & $20\times20$ & $23.190\pm0.054$ & $23.472\pm0.083$ & $24.365\pm0.050$ & $+0.893\pm0.116$ \\
\midrule
Beauty & $10\times10$ & $13.654\pm0.219$ & $13.952\pm0.205$ & $14.247\pm0.134$ & $+0.296\pm0.072$ \\
Sports & $10\times10$ & $8.433\pm0.051$ & $8.543\pm0.061$ & $8.431\pm0.058$ & $-0.112\pm0.015$ \\
Toys & $10\times10$ & $13.606\pm0.073$ & $13.868\pm0.097$ & $14.115\pm0.087$ & $+0.247\pm0.082$ \\
\bottomrule
\end{tabular}
\end{table*}

Refreshing the reference pool updates coverage supervision using the queries produced by CACO.
To separate this benefit from additional optimization, we compare three training schedules from the same frozen Base model: one epoch, two epochs with a fixed reference pool, and two epochs with a refresh after the first epoch.
All schedules use the same data splits and training seeds (42, 44, and 45), and inference applies a single adjustment network.
We report the mean and sample standard deviation across training seeds, averaging inference seeds 42 and 43 within each model.

Under the $20\times20$ retrieval setting used during training, refreshing the pool improves Recall@400 over the fixed-pool two-epoch control on Beauty, Sports, and Toys by 0.354--0.893 percentage points (Table~\ref{tab:appendix-refresh}).
The gains hold for every training seed on every dataset (Table~\ref{tab:appendix-refresh-seeds}), showing that the improvement comes from updated supervision rather than additional training alone.

The benefit is less consistent when transferring directly to $10\times10$ retrieval.
Mean Recall@100 improves on Beauty and Toys but decreases on Sports.
Reference refresh therefore strengthens coverage learning under this retrieval setting, while its benefit under a different query count and retrieval depth depends on the dataset.

\begin{table*}[t]
\centering
\caption{Test recall (\%) for each training seed, averaged over inference seeds 42 and 43. Both schedules train for two epochs; Refreshed updates the reference pool after the first epoch, while Fixed retains the original pool.}
\label{tab:appendix-refresh-seeds}
\small
\setlength{\tabcolsep}{13pt}
\begin{tabular}{llrrrr}
\toprule
& & \multicolumn{2}{c}{Recall@100} & \multicolumn{2}{c}{Recall@400} \\
\cmidrule(lr){3-4}\cmidrule(lr){5-6}
Dataset & Training seed & Fixed & Refreshed & Fixed & Refreshed \\
\midrule
Beauty & 42 & 13.791 & 14.130 & 22.694 & 23.441 \\
 & 44 & 13.882 & 14.218 & 22.707 & 23.525 \\
 & 45 & 14.182 & 14.394 & 23.288 & 23.852 \\
\addlinespace[2pt]
Sports & 42 & 8.582 & 8.453 & 16.886 & 17.372 \\
 & 44 & 8.575 & 8.475 & 17.160 & 17.522 \\
 & 45 & 8.472 & 8.366 & 17.255 & 17.470 \\
\addlinespace[2pt]
Toys & 42 & 13.757 & 14.066 & 23.452 & 24.420 \\
 & 44 & 13.909 & 14.063 & 23.563 & 24.323 \\
 & 45 & 13.937 & 14.215 & 23.400 & 24.351 \\
\bottomrule
\end{tabular}
\end{table*}

\subsection{Initial Overlap and Failure Cases}
\label{app:overlap-failures}
We examine how CACO's gains vary with the overlap among Base queries.
Test examples are assigned to low, middle, and high overlap groups using tertile boundaries estimated from Base $\operatorname{MPJ}$ on the validation set.
The lower and upper boundaries are $(0.0806,0.2642)$ for Beauty, $(0.1125,0.2167)$ for Sports, and $(0.0738,0.2240)$ for Toys.
Base and CACO are compared on the same examples with paired inference seeds.

Table~\ref{tab:appendix-overlap-strata} shows positive recall gains in every group, with the largest absolute improvements in the high-overlap group on all three datasets.
This pattern is consistent with query coordination being most beneficial when the initial retrieval sets are redundant, although the grouping alone does not establish causality.
The gained targets outnumber the lost targets in every group, explaining the overall recall improvement.

Broader coverage does not guarantee that every target is retained.
In Table~\ref{tab:appendix-failure-examples}, CACO returns more distinct items and lowers $\operatorname{MPJ}$ but loses a target retrieved by Base under both inference seeds.
For each dataset, we select the first such example in test-set order, using a loss under either seed only if no example fails under both.
These cases complement the successes in Appendix~\ref{sec:appendix-cases} and illustrate the distinction between improving aggregate coverage and preserving each individual prediction.
\begin{table*}[t]
\centering
\caption{Recall and coverage by initial Base overlap under $20\times20$ retrieval. Models use training seed 42 and inference seeds 42 and 43. Groups use validation-set Base MPJ averaged over inference seeds; $n$ is the number of test examples. Gained and Lost count targets newly retrieved or missed by CACO relative to Base, averaged over inference seeds. $\Delta$ is the recall gain in percentage points.}
\label{tab:appendix-overlap-strata}
\small
\setlength{\tabcolsep}{6pt}
\begin{tabular}{llrrrrrrrr}
\toprule
Dataset & Overlap group & $n$ & \multicolumn{2}{c}{Recall@400 (\%)} & $\Delta$ (pp) & \multicolumn{2}{c}{Target changes} & \multicolumn{2}{c}{MPJ} \\
\cmidrule(lr){4-5}\cmidrule(lr){7-8}\cmidrule(lr){9-10}
 & & & Base & CACO & & Gained & Lost & Base & CACO \\
\midrule
Beauty & All & 22363 & 18.307 & 23.666 & +5.359 & 1417.5 & 219.0 & 0.238 & 0.082 \\
Beauty & Low & 8193 & 8.861 & 9.752 & +0.891 & 149.5 & 76.5 & 0.048 & 0.029 \\
Beauty & Middle & 7146 & 14.281 & 17.024 & +2.743 & 309.0 & 113.0 & 0.148 & 0.038 \\
Beauty & High & 7024 & 33.421 & 46.654 & +13.233 & 959.0 & 29.5 & 0.550 & 0.189 \\
Sports & All & 35598 & 13.707 & 17.434 & +3.726 & 1760.0 & 433.5 & 0.184 & 0.050 \\
Sports & Low & 13289 & 11.020 & 12.435 & +1.415 & 363.5 & 175.5 & 0.072 & 0.029 \\
Sports & Middle & 11751 & 13.726 & 17.254 & +3.527 & 575.0 & 160.5 & 0.159 & 0.037 \\
Sports & High & 10558 & 17.068 & 23.925 & +6.857 & 821.5 & 97.5 & 0.355 & 0.090 \\
Toys & All & 19412 & 19.176 & 24.400 & +5.224 & 1157.5 & 143.5 & 0.197 & 0.062 \\
Toys & Low & 7059 & 10.200 & 11.262 & +1.062 & 128.0 & 53.0 & 0.045 & 0.026 \\
Toys & Middle & 6131 & 16.792 & 19.499 & +2.708 & 240.0 & 74.0 & 0.131 & 0.040 \\
Toys & High & 6222 & 31.710 & 44.134 & +12.424 & 789.5 & 16.5 & 0.433 & 0.124 \\
\bottomrule
\end{tabular}
\end{table*}

\begin{table*}[t]
\centering
\caption{Selected failure cases under $20\times20$ retrieval, shown for both inference seeds. CACO retrieves more distinct items with lower MPJ but misses the target retrieved by Base. Target hit is 1 for a hit and 0 for a miss.}
\label{tab:appendix-failure-examples}
\small
\setlength{\tabcolsep}{6pt}
\begin{tabular}{llrrrrrrr}
\toprule
Example & Dataset & Inference seed & \multicolumn{2}{c}{Target hit} & \multicolumn{2}{c}{Distinct items} & \multicolumn{2}{c}{MPJ} \\
\cmidrule(lr){4-5}\cmidrule(lr){6-7}\cmidrule(lr){8-9}
 & & & Base & CACO & Base & CACO & Base & CACO \\
\midrule
Example 1 & Beauty & 42 & 1 & 0 & 210 & 273 & 0.093 & 0.030 \\
Example 1 & Beauty & 43 & 1 & 0 & 215 & 289 & 0.108 & 0.028 \\
Example 2 & Sports & 42 & 1 & 0 & 175 & 214 & 0.211 & 0.077 \\
Example 2 & Sports & 43 & 1 & 0 & 179 & 223 & 0.186 & 0.070 \\
Example 3 & Toys & 42 & 1 & 0 & 264 & 289 & 0.038 & 0.027 \\
Example 3 & Toys & 43 & 1 & 0 & 272 & 299 & 0.030 & 0.022 \\
\bottomrule
\end{tabular}
\end{table*}

\subsection{Retrieval Efficiency and Budget Allocation}
\label{app:efficiency}
We examine the quality--speed trade-off at the common $L=K=20$ setting,
comparing Recall@400 with end-to-end inference time per sample.

All methods use a checkpoint trained with seed 42, with Recall averaged over
inference seeds 42 and 43 and time over five full test-set evaluations.
Measurements use H100 GPUs, batch size 32, four CPU threads per worker,
disjoint CPU affinity, and passive OpenMP waiting.
End-to-end time includes host-to-device transfer, query generation,
history-filtered exact cosine retrieval, device-to-host transfer, and
deduplication; loading, warmup, and offline preparation are excluded.
Discrete generators use the same SID-to-vector retrieval protocol as the
other shared-space comparisons.

\begin{figure*}[t]
\centering
\includegraphics[width=\textwidth]{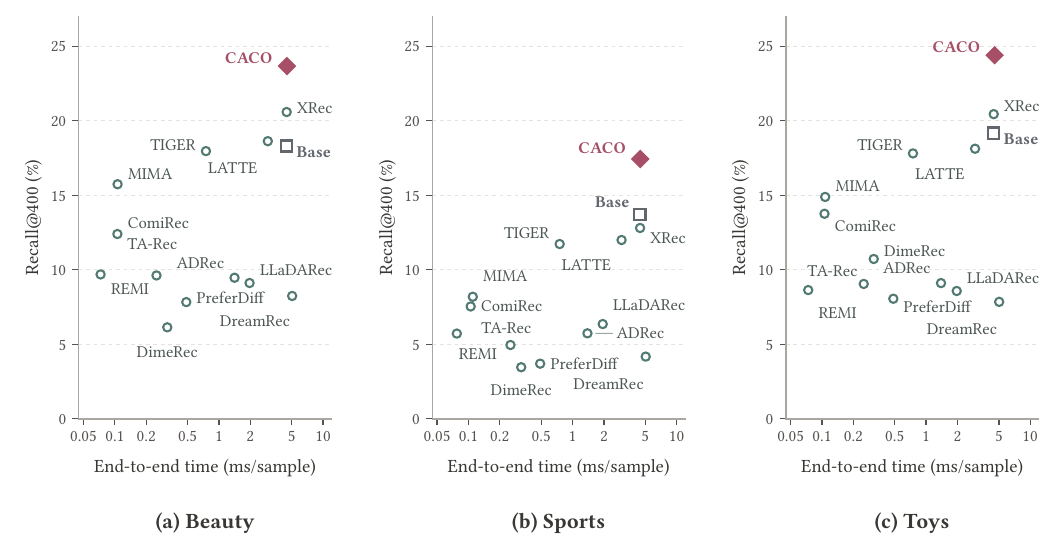}
\caption{Quality--speed trade-off under the shared $20\times20$ budget. Each point is one method; higher and further left is better. Each panel has its own logarithmic time axis and linear Recall axis, with identical limits across datasets. Rose diamonds highlight CACO, and gray squares denote Base. Horizontal bars show one timing standard deviation across five full test-set evaluations, generally smaller than the markers.}
\Description{Three separate scatter plots compare fourteen methods on Beauty, Sports, and Toys. Each has its own labelled axes and ticks: end-to-end milliseconds per sample on a logarithmic horizontal axis and Recall at 400 in percent on the vertical axis, using identical limits. Dataset labels (a) Beauty, (b) Sports, and (c) Toys appear below their respective panels. All methods are directly labelled. CACO is shown as rose diamonds, Base as gray squares, and other methods as open teal circles. CACO has the highest Recall in each panel at a time close to Base.}
\label{fig:appendix-recall-cost}
\end{figure*}

Figure~\ref{fig:appendix-recall-cost} shows that lightweight methods are
faster, while CACO achieves the highest Recall on all three datasets in
this comparison. Relative to Base, CACO raises Recall by 3.73--5.36
percentage points with an additional 0.03--0.10\,ms per sample
(0.7--2.3\%), preserving nearly the same inference cost.

Table~\ref{tab:appendix-budget-transfer} holds the retrieval budget at $B=400$ and compares $(L,K)=(10,40),(20,20),(40,10)$ without retraining CACO.
CACO improves recall, increases the number of distinct retrieved items, and reduces $\operatorname{MPJ}$ in all nine comparisons.
Both methods attain their highest recall with $10\times40$ among these allocations, showing that query count and retrieval depth remain consequential even under a fixed budget.
Table~\ref{tab:appendix-parameters} reports model sizes, including the frozen backbone in CACO; increasing the number of Base queries changes computation but adds no parameters.
\begin{table*}[t]
\centering
\caption{Query-count and depth allocation under a fixed retrieval budget of $B=400$. Each method uses a model trained with seed 42; results average inference seeds 42 and 43 on the full test set. CACO is trained with 20 queries and applied to all allocations without retraining. $\Delta$ is the Recall@400 gain in percentage points.}
\label{tab:appendix-budget-transfer}
\small
\setlength{\tabcolsep}{6pt}
\begin{tabular}{llrrrrrrr}
\toprule
Dataset & $L\times K$ & \multicolumn{2}{c}{Recall@400 (\%)} & $\Delta$ (pp) & \multicolumn{2}{c}{Distinct items} & \multicolumn{2}{c}{MPJ} \\
\cmidrule(lr){3-4}\cmidrule(lr){6-7}\cmidrule(lr){8-9}
 & & Base & CACO & & Base & CACO & Base & CACO \\
\midrule
Beauty & $10\times40$ & 21.171 & 25.231 & +4.060 & 197.2 & 267.6 & 0.260 & 0.101 \\
Beauty & $20\times20$ & 18.307 & 23.666 & +5.359 & 164.7 & 236.9 & 0.238 & 0.082 \\
Beauty & $40\times10$ & 15.682 & 22.086 & +6.403 & 140.2 & 211.0 & 0.213 & 0.066 \\
Sports & $10\times40$ & 15.630 & 18.498 & +2.868 & 199.1 & 286.4 & 0.206 & 0.063 \\
Sports & $20\times20$ & 13.707 & 17.434 & +3.726 & 163.9 & 255.2 & 0.184 & 0.050 \\
Sports & $40\times10$ & 11.908 & 16.175 & +4.267 & 137.4 & 228.3 & 0.166 & 0.039 \\
Toys & $10\times40$ & 21.940 & 25.811 & +3.871 & 200.9 & 276.7 & 0.226 & 0.081 \\
Toys & $20\times20$ & 19.176 & 24.400 & +5.224 & 169.1 & 248.4 & 0.197 & 0.062 \\
Toys & $40\times10$ & 16.971 & 23.120 & +6.148 & 144.4 & 224.5 & 0.171 & 0.047 \\
\bottomrule
\end{tabular}
\end{table*}

\begin{table*}[t]
\centering
\caption{Model parameters (millions) across the eight public benchmarks and the in-house industrial dataset. CACO includes the frozen backbone and one adjustment module. Frozen model components are counted; shared item embeddings and retrieval workspace are excluded. A dash denotes a method not included in the industrial comparison.}
\label{tab:appendix-parameters}
\small
\setlength{\tabcolsep}{4pt}
\begin{tabular*}{\textwidth}{@{\extracolsep{\fill}}l*{9}{r}@{}}
\toprule
Method & Beauty & Sports & Toys & Steam & Douban & Instruments & Games & Movies & Industrial \\
\midrule
Base & 8.465 & 8.465 & 8.465 & 8.465 & 8.465 & 8.465 & 8.465 & 8.465 & 448.350 \\
CACO & 9.094 & 9.094 & 9.094 & 8.961 & 8.961 & 9.094 & 9.094 & 9.094 & 449.375 \\
ComiRec & 7.970 & 7.970 & 7.970 & 7.970 & 7.970 & 7.970 & 7.970 & 7.970 & 451.905 \\
REMI & 7.970 & 7.970 & 7.970 & 7.970 & 7.970 & 7.970 & 7.970 & 7.970 & \textemdash \\
MIMA & 8.074 & 8.074 & 8.074 & 8.074 & 8.074 & 8.074 & 8.074 & 8.074 & \textemdash \\
TIGER & 8.083 & 8.094 & 8.088 & 8.080 & 8.077 & 8.081 & 8.098 & 8.221 & 450.827 \\
LATTE & 8.085 & 8.097 & 8.090 & 8.082 & 8.079 & 8.083 & 8.100 & 8.223 & 450.836 \\
LLaDARec & 8.708 & 8.708 & 8.708 & 8.214 & 8.214 & 8.708 & 8.708 & 8.708 & \textemdash \\
DreamRec & 8.049 & 8.049 & 8.049 & 8.049 & 8.049 & 8.049 & 8.049 & 8.049 & 453.869 \\
DimeRec & 7.959 & 7.959 & 7.959 & 7.959 & 7.959 & 7.959 & 7.959 & 7.959 & 450.186 \\
PreferDiff & 8.049 & 8.049 & 8.049 & 8.049 & 8.049 & 8.049 & 8.049 & 8.049 & \textemdash \\
TA-Rec & 8.164 & 8.164 & 8.164 & 8.049 & 8.049 & 8.164 & 8.164 & 8.164 & \textemdash \\
ADRec & 8.031 & 8.031 & 8.031 & 8.031 & 8.031 & 8.031 & 8.031 & 8.031 & \textemdash \\
X-Rec & 8.597 & 8.597 & 8.597 & 8.597 & 8.597 & 8.597 & 8.597 & 8.597 & 456.743 \\
\bottomrule
\end{tabular*}
\end{table*}

\subsection{Retrieval-Depth Expansion on the Industrial Dataset}
\label{app:industrial-depth}
\begin{figure*}[t]
\input{appendix_closing_layout_20261106_v2/depth_figure}
\end{figure*}
To distinguish broader query coverage from a larger output set, we compare depth expansion with and without a 400-item candidate cap (Figure~\ref{fig:appendix-industrial-depth}).
On the first partition of the industrial dataset, we keep each model's 20 queries fixed and vary CPU ANN depth from 20 to 400, using the same retrieval index and full test set.
Recall averages two inference seeds for one trained model per method; CPU ANN time averages five passes at batch size 32 and excludes query generation.
ComiRec uses its pretrained model, while TIGER retrieves through semantic-ID vectors in the shared embedding space.

Deeper retrieval improves every method when all distinct candidates are retained, but the gains largely saturate once the output is capped at 400 items.
CACO achieves the highest recall at every measured depth in both settings.
Under the cap, it first reaches 20\% recall at depth 32, compared with 200 for Base, with CPU ANN costs of 0.58 and 1.37\,ms per example, respectively.
Thus, coordinating queries improves coverage under the same output limit, beyond the benefit of simply returning more candidates.

\section{Dataset Statistics}
\label{app:dataset-statistics}
\renewcommand{\dbltopfraction}{0.95}
\renewcommand{\dblfloatpagefraction}{0.60}
Table~\ref{tab:appendix-dataset-statistics} summarizes the eight public datasets and the in-house industrial dataset.
Public statistics are reported after the preprocessing described in Section~\ref{sec:experimental-setup}.
The public datasets span 19,412--657,203 users and 11,924--197,943 items.
Interactions count retained positive events before histories are truncated to 200 items, and items refer to the full retrieval catalog.
Each user contributes one validation target and one test target; earlier prefixes provide training examples.

The in-house industrial dataset consists of 24 sequential partitions, each with approximately 2.6 million training instances and 10,240 test instances.
These counts refer to prediction instances and may include repeated users.

\begin{table*}[tbp]
\captionsetup{skip=3pt}
\caption{Dataset sizes after preprocessing.}
\label{tab:appendix-dataset-statistics}
\centering
\small
\setlength{\tabcolsep}{3.5pt}
\renewcommand{\arraystretch}{1.02}
\begin{tabular}{lrrr}
\toprule
Dataset & Users & Items & Interactions \\
\midrule
Beauty & 22,363 & 12,101 & 198,502 \\
Sports & 35,598 & 18,357 & 296,337 \\
Toys & 19,412 & 11,924 & 167,597 \\
Steam & 281,210 & 11,961 & 3,484,694 \\
Douban & 23,786 & 12,591 & 2,755,547 \\
Instruments & 57,439 & 24,587 & 511,836 \\
Games & 94,762 & 25,612 & 814,586 \\
Movies & 657,203 & 197,943 & 7,441,129 \\
\midrule
Dataset & Partitions & Train/part. & Test/part. \\
\midrule
Industrial & 24 & $\approx 2.6$M & 10,240 \\
\bottomrule
\end{tabular}

\end{table*}

\end{document}